\documentclass{svproc}
\usepackage{amsfonts,color,epsfig,epstopdf}
\usepackage{footnote,multirow}
\usepackage{amsmath}
\usepackage{amssymb}
\usepackage[utf8]{inputenc} 

\usepackage[varg]{txfonts}
\usepackage{mathrsfs}

\newcommand{\braket}[1]{\langle {#1} \rangle }
\newcommand{\ket}[1]{|{#1} \rangle }
\newcommand{\bra}[1]{\langle {#1}|}

\usepackage{color}
\usepackage{hyperref}

\begin{document}
\mainmatter              
\title{Indirect method for nuclear  reactions and the role of the self energy}
\author{Gregory Potel\inst{1}}
\institute{Departamento de Física Aplicada III, Escuela Superior de Ingenieros, Universidad de Sevilla, Camino de los Descubrimientos, Sevilla, Spain,\\
\email{gpotel@us.es}}

\maketitle              

\section{Introduction}
When a nuclear species (e.g., a nucleon or a deuteron nucleus) propagating freely is made to collide with a target nucleus, its trajectory is modified by exchanging variable amounts of energy, mass, linear and angular momentum with the target, according to its interaction with the nuclear medium. By addressing this perturbation away from the free path, one hopes to learn something about the nature of the medium through which our probe propagates. This is the essence of the experimental use of nuclear reactions for the purpose of  gathering information about nuclear structure. In order to deal with the structure and the reaction aspects of a specific experiment on the same footing, it is therefore desirable to identify a theoretical construct that embodies the modification of the propagation of a particle in the medium with respect to the free case, and use it both for the determination of the nuclear spectrum (structure) and for the calculation of scattering observables (reaction). A  candidate for such an object is the self energy, and we will try in the present lectures to put it at the center stage in the formulation of scattering theory.

Let us be more specific, and focus on a nucleon $x$ colliding with some target nucleus $A$. In all generality, the description of the scattering process can be written in terms of a many-body Hamiltonian, which will include the kinetic energy of the nucleon, the intrinsic Hamiltonian of the nucleus $A$, and the interaction between the nucleon and all the nucleons of $A$. However, as we will explicitly see below, this many-body Schr\"odinger equation can be reduced to a single-particle one, in which the nucleon propagates in an effective potential that includes the effects of the interaction with the medium. This effective potential is the self energy of the nucleon in the medium, and it is a complex, non-local, and energy-dependent operator. The imaginary part of this operator is responsible for the absorption of the nucleon in the target nucleus, i.e., the loss of flux with respect to the entrance (elastic) channel. The  scattering wavefunction $\psi_0$ of the resulting ($x$ + $A$) two-body system  associated with the propagation in the elastic channel can then be written in terms of the self energy, by means of the integro-differential Schr\"odinger equation
\begin{align}\label{eq:320}
\left(E-T\right)\psi_0({\mathbf r_{xA}})-\int  \Sigma'(\mathbf r_{xA},\mathbf r'_{xA},E)\,\psi_0({\mathbf r'_{xA}})\,d^3r'_{xA}=0,
\end{align}
where $E$ is the energy in the center of mass frame, $T$ is the kinetic energy operator acting on the $\mathbf r_{xA}$ relative coordinate\footnote{For simplicity, we omit the explicit reference to the spin variables.} of the nucleon with respect to the nucleus $A$, and $ \Sigma'$ is the self energy (with typical units in nuclear physics applications $[\Sigma']$=Mev$\cdot$fm$^{-3}$, i.e., a volume energy density), which we will also refer to as the optical potential. As we noted above, it is complex, energy-dependent and non-local, as a consequence of the coupling with the intrinsic degrees of freedom of the composite $(x+A)$ system (see Sect. \ref{SOP}). Non locality implies that ($\ref{eq:320}$) is an integro-differential equation, and the value of the associated eigenfunction $\psi_0$ at a given point $\mathbf r_{xA}$ depends on the values of the wavefunction at all other points $\mathbf r'_{xA}$. In order to clear up the notation, in what follows we will denote the action of the self energy (or any other non local operator) on a function $f(\mathbf r)$ of the space coordinates as
\begin{align}\label{eq:321}
\int  \Sigma'(\mathbf r,\mathbf r',E)\,f({\mathbf r'})\, d^3 r'\equiv\Sigma(\mathbf r,\mathbf r',E)\,f({\mathbf r}),
\end{align}
so that Eq. (\ref{eq:320}) can be written as
\begin{align}\label{eq:322}
\left(E-T-\Sigma(\mathbf r_{xA},\mathbf r'_{xA},E)\right)\,\psi_0({\mathbf r_{xA}})=0,
\end{align}
Let us make now an important point. The self energy $\Sigma$ (or $\Sigma'$) can, in principle, be derived explicitly and with arbitrary accuracy from the many-body Hamiltonian making use of well defined quantum many-body techniques \cite{dickhoffManyBodyTheory2005}. In this sense, solving Eq. (\ref{eq:320}) is equivalent to solving the many-body Schr\"odinger equation, and obtaining $\Sigma$ is, in general, equivalently hard (see Sect. \ref{SOP}). In the same way in which there are a number of techniques to solve the many-body Schr\"odinger equation, these methods can also be used to obtain the self energy.   We will stress in these lectures the use of nuclear field theory (NFT, \cite{bortignonNuclearFieldTheory1977,besNuclearFieldTheory1983}) in order to highlight the role that elementary modes of excitations (single-(quasi) particle states, collective pairing and surface vibrations, etc.) play both in structure and reactions.
\subsection{Indirect measurements of nuclear cross sections}
It is often the case that nuclear reactions that are important for societal applications or basic science are difficult to measure directly in existing facilities for accelerated beams. The difficulty might be associated with an exceedingly small cross section, as the ones obtained at beam energies well below the Coulomb barrier, or with the impossibility of devising a short-lived target, as is the case for neutron-induced reactions on unstable isotopes. The fruitful line of experimental research addressing this issue with alternative reactions (indirect measurements) has been developed in parallel to the theory needed to make the connection between the observed data and the desired cross sections of the  reaction under study \cite{escherCompoundNuclearReaction2012}. 

Within this context, the use of the self energy in Eq. (\ref{eq:322} )would provide the elastic and reaction cross sections associated with the  $x+A$ reaction, which can be computed from the asymptotic part of the wavefunction $\psi_0$ (see below). We will show in these lectures how the self energy of the $x-A$ system can be used to compute the cross section associated with the indirect measurement, which is defined as the inclusive measurement of a fragment $b$ produced in the reaction $a(=x+b)+A\to b+(x+A)$. 
Important examples of such indirect measurements are deuteron-induced reactions like $^{9}$Li$(d,p)^{10}$Li (see Sect. \ref{SVA}) and $^{40}$Ca$(d,p)^{41}$Ca (see Sect. \ref{SVB}).

The formalism to be presented in these lectures has been essentially introduced in the 80's \cite{Udagawa1986UT,Ichimura1985IAV,Ichimura1990IAV} and recently revived \cite{potelEstablishingTheoryDeuteroninduced2015,leiReexaminingClosedForm2015,carlsonElasticInelasticBreakup2015,potelCompleteTheoryPredicting2017}, in different contexts and variations. In order to be specific about this particular implementation,  we will call it  the Green's Function Transfer (GFT) formalism. The perturbative derivation presented here differs from earlier presentations, and it  might provide some different insights, including the possibility of estimating the validity of the spectator approximation (see Sect. \ref{SVIB}), and a more transparent connection with standard 2-body scattering theory techniques such as the $R$-matrix theory (see Sect. \ref{SVI}).  

\section{Elastic and inelastic scattering} 
\subsection{General formalism and definitions}   
Let us summarize in this Section some standard scattering theory results associated with the elastic scattering between two  nuclei $x$ and $A$. For a more detailed account, we refer the reader to some textbooks \cite{thompsonNuclearReactionsAstrophysics2009a,jacksonNuclearReactions1970,messiahMecaniqueQuantique1995,satchlerDirectNuclearReactions1983}, where a comprehensive exposition of quantum scattering theory in general, and nuclear reactions theory in particular, can be found. Within the context of the present lectures, we will treat $x$ as a structureless particle, while we will take the structure of both $A$ and $B\equiv A+x$ into full consideration. With this caveat in mind, the Hamiltonian of the system is\footnote{In the following, all the operators are assumed to be non-local, even if not indicated explicitly. Within this context, when it will be useful to specify the arguments of a given operator, we will just write them once for economy of notation, i.e., $A(\mathbf r)\equiv A(\mathbf r,\mathbf r')$. Note that there isn't any loss in generality in doing so, since a local operator is just a particular case of a non-local one: if $B(\mathbf r)$ is local, when acting on a general function $f(\mathbf r)$ it can be substituted with the non-local operator $B'(\mathbf r,\mathbf r')=B(\mathbf r)\delta(\mathbf r-\mathbf r')$. Then, $(Bf)(\mathbf r)=\int B'(\mathbf r,\mathbf r')f(\mathbf r')d^3r'=B(\mathbf r)\int \delta(\mathbf r-\mathbf r')f(\mathbf r')d^3r'=B(\mathbf r)f(\mathbf r)$.}
\begin{align}\label{eq301}
H=h_B(\mathbf r_{xA},\xi)=T_x+h_A(\xi)+V_{xA}(\mathbf r_{xA},\xi), 
\end{align}
where $\xi$ stands for all the spatial and spin coordinates needed to describe the microscopic structure of $A$, while $\mathbf r_{xA}$ is the relative coordinate of the $x$-$A$ system. For simplicity, we will ignore the intrinsic spins of $x$ and $A$.  The intrinsic Hamiltonians of nuclei $A$ and $B$ are $h_A$ and $h_B$, respectively, while $T_x$ is the kinetic energy associated to the  $x$-$A$ relative motion. The corresponding Schr\"odinger equation
\begin{align}
(E-H)\Psi(\mathbf r_{xA},\xi)=0, 
\end{align}
can be rewritten as
\begin{align}\label{eq311}
\left(E-T_x-h_A(\xi)\right)\Psi(\mathbf r_{xA},\xi)=V_{xA}(\mathbf r_{xA},\xi)\Psi(\mathbf r_{xA},\xi).
\end{align}
The previous expression suggests a formal solution in terms of the incident wave (defined with a suitable boundary condition specifying the beam direction, etc., see Eq. (\ref{eq314}) below) associated with the \emph{unperturbed} wavefunction,
\begin{align}\label{eq300}
 \left(E-T_{x}-h_{A}\right)\Psi_{0}=0.
\end{align}
and the inverse operator of the \emph{unperturbed} Hamiltonian
\begin{align}\label{eq310}
\mathbf G_{0}(\mathbf r_{xA},\xi,E)=\lim_{\eta\to0}\left(E-T_{x}-h_{A}+i\eta\right)^{-1},
\end{align}
which is the \emph{unperturbed} many-body Green's function, also called the propagator. The  small quantity $\eta$ should be made to vanish \emph{after} performing the inversion operation. Without its inclusion, the Green's function would be singular for values of the energy equal to the (real) eigenvalues of the Hamiltonian, and the inversion procedure would be ill defined. In addition, taking $\eta$ to be a \emph{positive} real vanishingly small number ensures that the second term after the equal sign in the equation below is proportional to an \emph{outgoing} spherical wave, thus enforcing the right asymptotic behaviour of the scattered wave (see, e.g., \cite{dickhoffManyBodyTheory2005,messiahMecaniqueQuantique1995}). It can be readily verified, by applying the operator $\left(E-T_{x}-h_{A}\right)$ to both sides of the equal sign of the equation below, that the wavefunction defined by the following Lippmann-Schwinger equation  
\begin{align}\label{eq400}
\Psi=\Psi_{0}+\mathbf G_{0}(E)V_{xA}\Psi
\end{align}
indeed satisfies Eq. (\ref{eq311}). This equation can also be expressed in an equivalent way\footnote{The equivalence between these two forms  is a standard scattering theory result, which can be obtained making use of the equation connecting $\mathbf G_{0}$ and $\mathbf G$ (Dyson's equation) (see, e.g., \cite{dickhoffManyBodyTheory2005}),
\begin{align}
\mathbf G=\mathbf G_{0}+\mathbf G_{0} V_{xA} \mathbf G=\mathbf G_{0}+\mathbf G V_{xA} \mathbf G_{0}.
\end{align}
},
\begin{align}\label{eq312}
\Psi=\Psi_{0}+\mathbf G(E)V_{xA}\Psi_{0}.
\end{align}
where 
\begin{align}\label{eq401}
\mathbf G(\mathbf r_{xA},\xi,E)=\lim_{\eta\to0}\left(E-T_{x}-h_{A}-V_{xA}+i\eta\right)^{-1}
\end{align}
is the \emph{total} many-body Green's function. The first term on the right hand side of Eqs. (\ref{eq400}) and (\ref{eq312}) is the incident wave, describing the incident channel imposed by our boundary condition, while the second term is the scattered wave.

 The unperturbed wavefunction solution of Eq. (\ref{eq300}) associated with standard scattering boundary conditions is
\begin{align}\label{eq314}
\Psi_{0}(\mathbf r_{xA},\xi)=\Phi_{0}(\xi)F (\mathbf k_{x},\mathbf r_{xA}),
\end{align}
where $F (\mathbf k_{x},\mathbf r_{xA})$ is a free incoming incident plane wave with momentum $\mathbf k_{x}$ along the $z$ axis\footnote{For the rather common case in which both $x$ and $A$ are charged, it is better to exclude from the definition of $V_{xA}$ the Coulomb interaction, and include it in the unperturbed Green's function
\begin{align}
\mathbf G_{0}=\left(E-T_{x}-h_{A}-\frac{e^{2}Z_{A}Z_{x}}{r_{xA}}\right)^{-1}.
\end{align}
In this case, $F (\mathbf k_{x},\mathbf r_{xA})$ would be a Coulomb function, without otherwise affecting the overall discussions in these lectures.}, and $\Phi_{0}$ is the ground state of the nucleus $A$ (see Eq. (\ref{eq303}) below).

\subsubsection{Elastic scattering}

We can expand the wavefunction describing the $x$-$A$ system in terms of the intrinsic states of $A$ as
\begin{align}\label{eq302}
    \Psi(\mathbf r_{xA},\xi;E)=\sum_i\Phi_i(\xi)\psi_i(\mathbf r_{xA}),
\end{align}
where $\left\{\psi_i(\mathbf r_{xA})\right \}$ is a complete set of orthogonal (not necessarily normalized) single-particle \textit{channel} wavefunctions, and $\left\{\Phi_i(\xi)\right \}$ is also a complete set of eigenfunctions of $h_A$,
\begin{align}\label{eq303}
   \left(\epsilon_i-h_A(\xi)\right)\Phi_i(\xi)=0.
\end{align}
In order to obtain an equation for the elastic  ($i=0$) channel, we project the first Eq. (\ref{eq300}) on $\Phi_{0}$, obtaining a Lippmann-Schwinger equation for the one-body elastic channel wavefunction,
\begin{align}\label{eq402}
\psi_0(\mathbf r_{xA})=F+\bra{\Phi_{0}}\mathbf G(E)V_{xA}\ket{\Phi_{0}}F.
\end{align}
\subsubsection{Inelastic scattering}
Eq. (\ref{eq402}) can be easily generalized to the description of inelastic scattering, where the nucleus $A$ has been excited to a state $i$ of its spectrum,
\begin{align}\label{eq403}
\psi_i(\mathbf r_{xA})=\bra{\Phi_{i}}\mathbf G(E)V_{xA}\ket{\Phi_{0}}F.
\end{align}
Note that in this case the free, unscattered wafefunction $F$ does not appear, testifying to the fact that, as part of our asymptotic boundary condition, only the ground state of $A$ is in the incident channel. The population of the inelastic $(i\neq0)$ channels is entirely due to the scattering process driven by the interaction $V_{xA}$.
\subsubsection{Cross section and $T$-matrix}
The cross section is associated with the asymptotic form of $\psi_{i}$, which describes the observed system far away from the interaction region, where the detectors are located. The asymptotic form of the elastic and inelastic channel wavefunctions can be obtained from Eqs. (\ref{eq402}) and (\ref{eq403}), respectively, by making use of the asymptotic form of the Green's function (see Eq. (\ref{eq410}) below),
\begin{align}\label{eq319}
\nonumber \psi_{i}(r_{xA}\to\infty)&=F(r_{xA}\to\infty)\delta_{i0}+\bra{\Phi_{i}}\mathbf G(r_{xA}\to\infty)V_{xA}\ket{\Phi_0} F\\
\nonumber &=F(r_{xA}\to\infty)\delta_{i0}+O(\mathbf k_{i},\mathbf r_{xA})\bra{\psi_{i}\Phi_{i}}V_{xA}\ket{\Phi_0 F}\\
&=F(r_{xA}\to\infty)\delta_{i0}+O(\mathbf k_{i},\mathbf r_{xA})T_{i0},
\end{align} 
where $O(\mathbf k_{i},\mathbf r_{xA})$ is an outgoing  wave with momentum $k_{i}=\sqrt{2\mu(E-\epsilon_{i})}/\hbar$, the amplitude
\begin{align}\label{eq404}
T_{i0}=\bra{\psi_{i}\Phi_{i}}V_{xA}\ket{\Phi_0 F}
\end{align}
is an element of the $T$-matrix,  and we have used the asymptotic form of the Green's function,
\begin{align}\label{eq410}
\bra{\Phi_i}\mathbf G(r_{xA}\to\infty,r'_{xA},\xi,\xi')\equiv\int\Phi_i(\xi)^*\,\mathbf G(r_{xA}\to\infty,r'_{xA},\xi,\xi')\,d\xi=O(\mathbf k_{i},\mathbf r_{xA})\Phi_{i}(\xi')\psi_{i}(\mathbf r'_{xA}).
\end{align}
The differential cross section can be expressed in terms of the $T$-matrix (see, e.g., \cite{thompsonNuclearReactionsAstrophysics2009a,jacksonNuclearReactions1970}),
\begin{align}\label{eq435}
\frac{d\sigma_{i0}}{d\Omega}=\frac{\mu^{2}k_{i}}{4\pi^{2}\hbar^{4} k_{0}}|T_{i0}|^{2}.
\end{align}
\subsubsection{$R$-matrix parametrization of the $T$-matrix}
In the context of 2-body quantum scattering theory, cross sections can always be calculated in terms of the phase shift existing between the free and scattered wavefunctions in the asymptotic region, i.e., sufficiently far away from the scattering center (see, e.g \cite{thompsonNuclearReactionsAstrophysics2009a,jacksonNuclearReactions1970,messiahMecaniqueQuantique1995}). This phase shift uniquely determines the asymptotic behavior of the scattered wavefunction, and it can in turn be inferred from the (inverse) logarithmic derivative of the  wavefunction at some arbitrary radius in the asymptotic region. This quantity (divided by the radius at which it has been calculated) becomes a dimensionless matrix $R_{ij}(E)$ (the so-called $R$-matrix, dependent on the center of mass energy $E$) when we take into account the population of different reaction channels (labeled by the indexes $i,j$), each one of them associated with its own wavefunction. 

It can be shown (\cite{laneRMatrixTheoryNuclear1958,descouvemontRmatrixTheory2010}) that the energy-dependence of the $i,j$ element of the $R$-matrix can be exactly parametrized in terms of an infinite set of real, energy-independent parameters $\gamma_{ip},\gamma_{jp},E_p$  as
 \begin{align}
 R_{ij}(E)=\sum_{p}^\infty\frac{\gamma_{ip}\gamma_{jp}}{E_{p}-E}.
 \end{align}
This is known as the Wigner-Eisenbud parametrization. The $\gamma_{ip}$ are known as the reduced partial widths (with dimensions of $[\gamma_{ip}]=E^{1/2}$) , and the $E_p$ are the corresponding poles of the $R$-matrix.

 Although these quantities can be just considered as parameters to be adjusted in order to fit the observed experimental cross section (see below), they have a specific interpretation in terms of the so-called \emph{calculable} $R$-matrix formalism (\cite{laneRMatrixTheoryNuclear1958,descouvemontRmatrixTheory2010}).   The energies $E_p$ are the eigenvalues of the eigenvalue problem associated with the Hamiltonian of the system and \emph{specific boundary conditions at the radius $a$ where the $R$-matrix is calculated}, while the $\gamma_{ip}$ are related to the amplitude of the corresponding eigenfunction at this radius. A variety of boundary conditions can be chosen giving rise to different practical implementations of $R$-matrix theory, but they all have in common that the resulting spectrum is discrete (see e.g. \cite{descouvemontRmatrixTheory2010,blochFormulationUnifieeTheorie1957}). 
 
 Needless to say, these quantities depend on the choice of the $R$-matrix radius $a$ and the specific boundary conditions chosen. Because of the non-physical boundary contions implemented, the energies $E_p$ do not match the eigenvalues of the physical problem, which corresponds instead to the eigenvalue problem associated with the physical boundary conditions (exponentially decaying for bound states, oscillating for scattering states). Similarly, the energies $E_p$ do not correspond to the physical resonances of the system, which are instead associated with the poles of the $T$-matrix (see Eq. (\ref{eq332})).

Like any other asymptotic quantity, the $T$-matrix can then be  derived in terms of the $R$-matrix parameters, resulting in (\cite{laneRMatrixTheoryNuclear1958})
  \begin{align}\label{eq415}
 T_{ij}=\sqrt{P_{i}(E)P_{j}(E)}\sum_{pq}^\infty\gamma_{ip}\,[A^{-1}(E)]_{pq}\,\gamma_{jq},
 \end{align}
 with
  \begin{align}
 A_{pq}(E)=(E_{p}-E)\delta_{pq}-\sum_{c}\gamma_{cp}\gamma_{cq}(S_{c}(E)+iP_{c}(E)).
 \end{align}
 The energy dependence of the $T$-matrix is thus contained in the $(E-E_{p})$ term in the denominator, and in the penetrability ($P_{i}(E)$) and shift $(S_{i}(E))$ factors, which are known combinations of the Bessel functions and their derivatives (or Coulomb functions, if both particles associated with the reaction channel $i$ are charged) \cite{descouvemontRmatrixTheory2010} .  The physical resonances (i.e., sharp structures in the strength function $d\sigma(E)/dE$ as a function of the energy) are the complex poles $\mathcal E_{p}$ of the $T$-matrix, which can be identified as the  solutions of the implicit equation
 \begin{align}\label{eq332}
 A_{pp}(\mathcal E_{p})=0,
 \end{align} 
 for every $p$. Since these are physical, experimentally observable quantities, they do not depend on the choice of the $R$-matrix radius or the boundary conditions used to calculate the $R$-matrix. The imaginary part of $\mathcal E_{p}$ is associated with the width of the resonance, while its real part is the resonance energy.
 
 Any practical implementation of the so-called \emph{phenomenological $R$-matrix} usually consists in fitting a finite number of energies $E_p$ and partial widths $\gamma_{ip}$ to an experimental excitation function $d\sigma(E)/dE$, which is proportional to the modulus square of the $T$-matrix (\ref{eq415} )(see Eq. (\ref{eq435})). Since the energy dependence of the $T$-matrix is then known explicitly, the cross section can be calculated at any energy, including the low energies of astrophysical interest, where direct measurements are often not feasible (see e.g. \cite{deBoer2017RMP,Formicola2004PLB}). 
\subsubsection{Born series and the Distorted Wave Born Approximation}\label{DWBA}
The expression (\ref{eq400}) can be used to express the wavefunction  as a perturbative expansion in  terms of powers of the potential $V_{xA}$, known as the \emph{Born series},
 \begin{align}
\Psi=\Psi_{0}+\mathbf G_{0}(E)V_{xA}\left(\Psi_0+\mathbf G_{0}(E)V_{xA}(\Psi_0+\dots\right)).
\end{align}
The first order term of this expansion,
  \begin{align}
\Psi\approx\Psi_{0}+\mathbf G_{0}(E)V_{xA}\Psi_0
\end{align}
is the first order \emph{Plane Wave Born Approximation}. However, this is often not the most convenient way to express the perturbation expansion for the wavefunction. Instead, one often defines an arbitrary auxiliary potential $U_0(\mathbf r_{xA})$ and, instead of Eq. (\ref{eq311}), has
\begin{align}\label{eq316}
\left(E-T_x-h_A(\xi)-U_0(\mathbf r_{xA})\right)\Psi(\mathbf r_{xA},\xi)=\left(V_{xA}(\mathbf r_{xA},\xi)-U_0(\mathbf r_{xA})\right)\Psi(\mathbf r_{xA},\xi).
\end{align}
The solution to the above equation can now be expressed in the two equivalent following ways
\begin{align}\label{eq4011}
\nonumber \Psi&=\tilde \Psi_{0}+\tilde{\mathbf G}_{0}(E)\left(V_{xA}(\mathbf r_{xA},\xi)-U_0(\mathbf r_{xA})\right)\Psi;\\
\Psi&=\tilde \Psi_{0}+\mathbf G(E)\left(V_{xA}(\mathbf r_{xA},\xi)-U_0(\mathbf r_{xA})\right)\tilde\Psi_0,
\end{align}
where
\begin{align}\label{eq416}
\tilde{\mathbf G}_{0}=\lim_{\eta\to0}\left(E-T_{x}-h_{A}-U_0+i\eta\right)^{-1},
\end{align}
and 
\begin{align}\label{eq320}
\tilde \Psi_{0}(\mathbf r_{xA},\xi)=\Phi_{0}(\xi)\tilde F (\mathbf r_{xA}).
\end{align}
Now, $\tilde F$ is the so-called \emph{distorted wave}, and is the solution of
\begin{align}\label{eq417}
\left(E-T_{x}-h_{A}-U_0\right)\tilde F=0.
\end{align}
The first order approximation to the associated power series,
  \begin{align}
\Psi\approx\tilde \Psi_{0}+\tilde{\mathbf G}_{0}(E)\left(V_{xA}(\mathbf r_{xA},\xi)-U_0(\mathbf r_{xA})\right)\tilde \Psi_0
\end{align}
is know as the first order \emph{Distorted Wave Born Approximation} (DWBA), and it is widely used in practice (see e.g. \cite{thompsonNuclearReactionsAstrophysics2009a,jacksonNuclearReactions1970}). The freedom in choosing the auxiliary potential $U_0$ can be used to pick one which makes the matrix elements of $\left(V_{xA}(\mathbf r_{xA},\xi)-U_0(\mathbf r_{xA})\right)$ relevant for the calculation of $\Psi$ as small as possible, in order to speed up the convergence of the power series. 

In what follows, all the derivations made using $\mathbf G_0$ and $F$ remain valid if we make the substitutions
  \begin{align}\label{eq321}
\mathbf G_{0}\to \tilde{\mathbf G}_{0};\quad F\to\tilde F;\quad V_{xA}\to V_{xA}-U_0,
\end{align}
even when not stated explicitly. The shift of the zero-order many-body wavefunction implied in going from $\Psi_0$ to $\tilde\Psi_0$ is  common practice in actual nuclear reaction calculations. 
\subsection{The optical potential}\label{SOP}
Our ability to calculate elastic and inelastic cross sections according to Eqs. (\ref{eq402}) and (\ref{eq403}) rely on being able to obtain matrix elements of the Green's function $\mathbf G$, possibly  within some reasonable approximations.  We now show how the knowledge of the optical potential (which we show to be equivalent to the self energy defined in Eq. (\ref{eq:322})) provides a practical way to address this problem. It is essentially a tautology to say that the calculation of the optical potential allows for the calculation of the elastic scattering wavefunction, but we will use this explicit connection in a less trivial context in Section \ref{Sec2}.

Let us start by looking for a one-body Schr\"odinger equation for  the wavefunction $\psi_0$ associated to elastic scattering for an energy $E$. Projecting the many-body Schr\"odinger equation
\begin{align}\label{eq304}
   \left(E-H\right)\Psi=0
\end{align}
on the states $\Phi_i$, one obtains a set of coupled equations for the one-body states $\psi_i$ (see Eq. (\ref{eq302})),
\begin{align}\label{eq305}
  \nonumber  &\left(\vphantom{\sum f}E_0-T_x-V_{00}(\mathbf r_{xA})\right)\psi_0(\mathbf r_{xA})=\sum_{i\neq0}V_{0i}(\mathbf r_{xA})\,\psi_i(\mathbf r_{xA})\\
   &\left(\vphantom{\sum f}E_i-T_x-V_{ii}(\mathbf r_{xA})\right)\psi_i(\mathbf r_{xA})=V_{i0}(\mathbf r_{xA})\,\psi_0(\mathbf r_{xA})+\sum_{j\neq0,i}V_{ij}(\mathbf r_{xA})\,\psi_j(\mathbf r_{xA})\quad (i\neq0),
\end{align}
where the \emph{coupling potentials} are defined as
\begin{align}\label{eq306}
    V_{ij}(\mathbf r_{xA})=\int d\xi \,\Phi^*_i(\xi) \,V_{xA}(\mathbf r_{xA},\xi)\,\Phi_j(\xi),
\end{align}
and $E_i=E-\epsilon_i$.
The desired solution can be obtained within the Coupled Channels approach by solving numerically the set of coupled differential equations (\ref{eq305}) making use of some reasonable approximations \cite{satchlerDirectNuclearReactions1983,tamuraAnalysesScatteringNuclear1965,thompsonCoupledReactionChannels1988}. Alternatively, one can use the propagator (Green's function) \emph{restricted to the space of the excited states of the $x+A$ system}, 
\begin{align}\label{eq318}
    G^Q_{ij}(\mathbf r_{xA};E)=\lim_{\eta\to0}\left(E-(T_x+\epsilon_j)\,\delta_{ij}-V_{ij}+i\eta\right)^{-1}\quad (i,j\neq0)
\end{align}
to solve for the $\psi_i$'s in terms of $\psi_0$,
\begin{align}\label{eq307}
    \psi_i=\sum_{j\neq0}  G^Q_{ij}V_{j0}\,\psi_0,
\end{align} 
and substitute in the first equation of (\ref{eq305}),
\begin{align}\label{eq308}
    \left(\vphantom{\sum f}E_0-T_x-\mathcal V(\mathbf r_{xA},E)\right)\psi_0(\mathbf r_{xA})=0.
\end{align}
The non-local, complex, and energy dependent operator
\begin{align}\label{eq309}
    \mathcal V(\mathbf r_{xA},E)=V_{00}+\sum_{ij}V_{0i}G^Q_{ij}(E)V_{j0}
\end{align}
is the optical potential. The superscript $^Q$ in the Green's function indicates that it is restricted to the portion of the Hilbert space spanned by the excited states, excluding the ground state $\psi_0$. In other words,
\begin{align}\label{eq309b}
G^Q\psi_0=0.
\end{align}
Following the standard notation of \cite{feshbachUnifiedTheoryNuclear1962}, we will call this subspace the $Q$ space.

 The optical potential is seen to be equivalent to the self energy by comparing Eq. (\ref{eq308}) with Eq. (\ref{eq:322}). The discussion in the present Section also illustrates how the one-body Eq. (\ref{eq308}) is completely equivalent to solving the coupled equations (\ref{eq305}) for the elastic channel, and, therefore, to diagonalizing the Hamiltonian, as mentioned in the Introduction.
     
  The direct numerical implementation of Eq. (\ref{eq309}) in a truncated $Q$ model space is not the only way to address the calculation of the optical potential. For example, in the NFT approach to nuclear structure and reactions (see, e.g., \cite{r.a.brogliaUnifiedDescriptionStructure2016,potelNuclearCooperPair2021}), the self energy is calculated making use of Feynman diagrams, which provide a systematic way to take into account the coupling of single-particle motion with collective surface and pairing vibrations, which are calculated making use of the Quasi-Particle Random Phase Approximation (QRPA) wavefunctions. In this case, the self energy is obtained as a perturbative expansion in terms of the particle-vibration coupling vertices, which provide a specific model for the couplings $V_{0i}$. As a result of this perturbative diagonalization,  the  self energy obtained is non-local, complex, and energy dependent, as expected from Eq. (\ref{eq309}). In Sect. \ref{SVA} we provide a specific application addressing the $^9$Li$(d,p)$ reaction making use of NFT.  We refer  to the contribution of F. Barranco \emph{et al.} to this volume for more details on the NFT approach.

 The energy dependence of the optical potential (\ref{eq309}) is contained in the virtual propagation in the $Q$ space, encoded in the propagator $G^Q(E)$. It is thus a consequence of the connection of the ground state with the excited states described by the couplings $V_{0i}$. In order to gain some insight on the nature of its imaginary  part, let us consider  the matrix element of the propagator in a  eigenstate $\ket{\mu}$ of the Hamiltonian restricted to the $Q$ space,
 \begin{align}\label{eq322}
\left(\varepsilon_\mu-T_x-V(\mathbf r_{xA})\right)\ket{\mu}=0,
\end{align}
then
 \begin{align}\label{eq324}
G^Q_\mu(E)=\braket{\mu|G^Q(E)|\mu}=\lim_{\eta\to0}\frac{1}{E-\varepsilon_\mu+i\eta}.
\end{align}
Let us extract the imaginary part of this matrix element,
\begin{align}
  \nonumber \text{Im}\,G^Q_\mu(E)=\text{Im}&\left(\lim_{\eta\to0}\frac{1}{E-\varepsilon_\mu+i\eta}\right)=\text{Im}\left(\lim_{\eta\to0}\frac{E-\varepsilon_\mu-i\eta}{(E-\varepsilon_\mu+i\eta)(E-\varepsilon_\mu-i\eta)}\right)\\
  &=\text{Im}\left(\lim_{\eta\to0}\frac{E-\varepsilon_\mu-i\eta}{(E-\varepsilon_\mu)^2+\eta^2}\right)=-\pi\delta(E-\varepsilon_\mu),
\end{align} 
 where we have used the  representation of the Dirac delta function,
 \begin{align}\delta(x)=\lim_{\eta\to0}\frac{1}{\pi}\frac{\eta}{x^2+\eta^2}.
 \end{align}
 In keeping with the fact that the energy dependence of the optical potential is entirely contained in the propagator $G^Q(E)$, the above result illustrates the fact that the optical potential becomes complex whenever the energy $E$ matches one of the eigenvalues $\varepsilon_\mu$ of the Hamiltonian in the $Q$ space. As a consequence, the Hamiltonian describing the dynamics of the elastic channel is not Hermitian, and the flux in this channel is not conserved. Since it can be shown that this imaginary part is always negative (see, e.g., \cite{feshbachUnifiedTheoryNuclear1958}, where a more rigorous derivation of the above results can also be found), this flux non-conservation consists in a loss (rather than an increase) of flux in the elastic channel, which leaks into the $Q$ space. This feeding of the excited states of the $x+A$ system is what we call a reaction event, which, of course, is an energy-conserving process.  The explicit connection of the imaginary part of the optical poterntial with the reaction cross section is expressed by Eq. (\ref{eq407}) below.

  The elastic scattering wavefunction  can then be written as the solution of the Lippmann-Schwinger equation \emph{involving exclusively single-particle operators},
\begin{align}\label{eq313}
   \psi_0(\mathbf r_{xA})=F+G_{0}(E_{0})\,\mathcal V(E)\,\psi_0,
\end{align}
where
\begin{align}\label{eq315}
G_{0}(\mathbf r_{xA},E)=\left(E-T_x\right)^{-1}
\end{align}
is the free \emph{single-particle} Green's function.  
Eq. (\ref{eq313}) can be written in a different exact form (see Eq. (\ref{eq400})),
\begin{align}\label{eq317}
    \psi_0=F+G(E_{0})\,\mathcal V(E)\,F,
\end{align}
where $G$ is the  \emph{single-particle} Green's function,
\begin{align}\label{eq413}
G(\mathbf r_{xA},E)=\left(E-T_x-\mathcal V(E)\right)^{-1}.
\end{align}
We can now use (\ref{eq402}) to identify
\begin{align}\label{eq406}
\bra{\Phi_{0}}\mathbf G(E)V_{xA}\ket{\Phi_{0}}=G(E_{0})\,\mathcal V(E).
\end{align}

Another standard result of scattering theory is that the total reaction cross section, corresponding to the sum over  all energetically allowed excitations of the $x$-$A$ system, can be obtained as the expectation value of the imaginary part of the optical potential over the elastic wave function (see, e.g., \cite{thompsonNuclearReactionsAstrophysics2009a,jacksonNuclearReactions1970,satchlerDirectNuclearReactions1983}),
\begin{align}\label{eq407}
\sigma_{R}=\frac{2\mu}{\hbar^2 k_{x}}\braket{\psi_{0}|\textrm{Im}\mathcal V
|\psi_{0}},
\end{align}
 where $\mu$ is the reduced mass of the $x-A$ system, and $k_{x}$ is the associated wave number in the center of mass frame.
  \section{Indirect measurements and GFT}\label{Sec2}
  \subsection{Introduction: effective 3-body problem and the spectator approximation}
\begin{figure}[h]
	\centerline{\includegraphics[width=15cm]{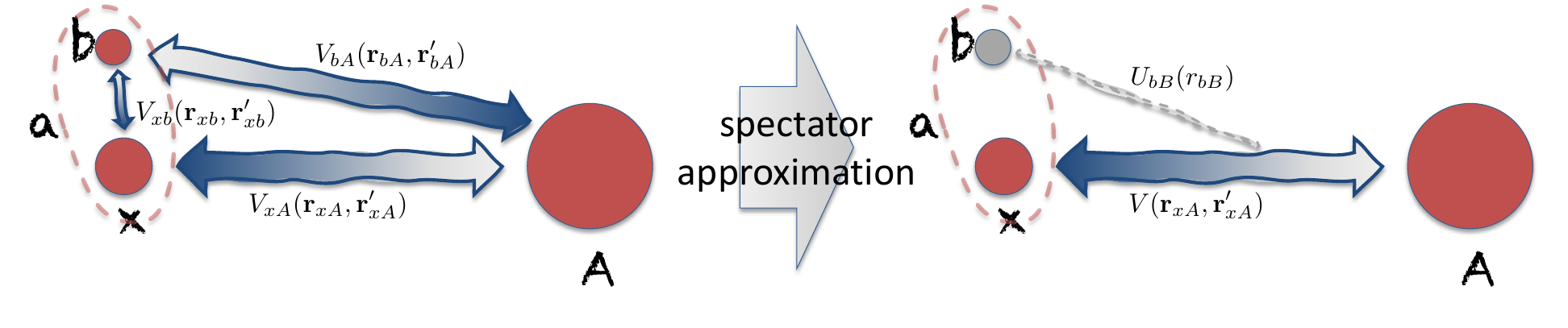}}
	\caption{Schematic representation of the inclusive reaction $a+ A\to b+(x+A)$, where the fragment $b$ is detected, and the nucleon $x$ is not. The interaction between $b$ and the rest of the system is approximated with an optical potential. The spectator approximation assumes that the interaction between $x$ and $A$ can be treated as a two-body problem, in which the fragment $b$ acts as a spectator.} 
	\label{fig1} 
\end{figure}

Let us now consider a composite system $a\equiv x+b$ impinging on the nucleus $A$ (see Fig. \ref{fig1}). We want to deal here with measurements in which the projectile $a$ breaks into its constituents $x$ and $b$, and only $b$ is detected with an energy $E_b$.  The fragment $b$ will also be assumed to be structureless, and its propagation will be described within the \textit{spectator approximation}. The  Hamiltonian can be written in two equivalent  representations,
\begin{align}\label{eq500}
    \nonumber &H=H_{B}=T_{x}+h_A(\xi)+V_{xA}(\mathbf r_{xA},\xi)+T_b+U_{bB}(\mathbf r_{bB})+\Delta V_{B}\\
    &=H_{A}=T_a+h_a(\mathbf r_{xb})+h_A(\xi)+V_{xA}(\mathbf r_{xA},\xi)+U_{bA}(\mathbf r_{bA})+\Delta V_{A},
\end{align}
where $h_a$ is the intrinsic Hamiltonian of the nucleus $a$, and we define
\begin{align}\label{eq328}
    \nonumber &\Delta V_{B}=V_{xb}(\mathbf r_{xb})+V_{bA}(\mathbf r_{bA},\xi)-U_{bB}(\mathbf r_{bB});\\
    &\Delta V_{A}=V_{bA}(\mathbf r_{bA},\xi)-U_{bA}(\mathbf r_{bA}).
\end{align}

Methods exist that deal with the effective three-body problem ($x+b+A$) in an essentially exact way, such as the Faddeev equations \cite{AltGrassbergerSandhas1967,Gloeckle1983,FaddeevMerkuriev1993}, and the Continuum Discretized Coupled Channels (CDCC) method \cite{Austern1987CDCC,Kamimura1986PTPS,Yahiro2012PTEP}. However, these methods are computationally expensive, and they are not always practical for the analysis of experimental data. The spectator approximation is a relatively economical way to reduce the three-body problem to two  two-body ones (($x+A$) and ($b+(x+A)$)), in which the fragment $b$ acts as a spectator. In addition, the role of the interaction between $x$ and $A$ (and, therefore, of the reaction corresponding to the \emph{direct} measurement we wish to study) in the experimental cross section for the observation of the fragment $b$ is cleanly highlighted.  
In the context of the spectator approximation  the interaction between the fragment $b$ and the rest of the system is approximated  with the optical potentials $U_{bB}$ and $U_{bA}$, so we assume $\Delta V_{A}\approx 0$, and $\Delta V_{B
}\approx 0$. Within this context, the coupling of the fragment $b$ to the intrinsic structure of the systems $A$ and $B$ is only  phenomenologically included in terms of an imaginary part of the potentials $U_{bB}$, $U_{bA}$. In Sect. \ref{SVIB} we will give a more precise meaning to the condition that the operators $\Delta V_{A}$ and $\Delta V_{B}$ are negligible.

\subsection{Elastic scattering and inclusive cross section} 
Let us write the Schr\"odinger equation for the total wavefunction $\Psi(\mathbf r_{xA},\mathbf r_{bA},\xi)$ assuming $\Delta V_{A}=0$,
\begin{align}\label{eq551}
\left(E-T_a-h_a(\mathbf r_{xb})-h_A(\xi)\right)\Psi=\left(V_{xA}(\mathbf r_{xA},\xi)+U_{bA}(\mathbf r_{bA})\right)\Psi,
\end{align}
which can be formally solved in terms of many-body Green's function,
\begin{align}\label{eq408}
\Psi=\Psi_{0}+\mathbf G_{I}(E)\left[V_{xA}(\mathbf r_{xA},\xi)+U_{bA}(\mathbf r_{bA})\right]\Psi_{0},
\end{align}
where the have defined the unperturbed wavefunction
\begin{align}\label{eq329}
\Psi_{0}=\phi_{a}(\mathbf r_{xb})\Phi_{0}(\xi)F (\mathbf k_{a},\mathbf r_{a}),
\end{align}
where $\mathbf r_{a}$ is the coordinate of the center of mass of the nucleus $a$, and $F(\mathbf k_{a},\mathbf r_{a})$ is a free incoming  wave (see previous section). The unperturbed wavefunction satisfies
\begin{align}
\left(E-T_a-h_a(\mathbf r_{xb})-h_A(\xi)\right)\Psi_{0}=\left(E-T_a-\epsilon^{a}_{0}-\epsilon^{A}_{0}\right)\Psi_{0}=\left(E_{0}-T_a\right)\Psi_{0}=0,
\end{align}
and 
\begin{align}
\left(\epsilon_0^{a}-h_a(\mathbf r_{xb})\right)\phi_{a}=0,\quad \left(\epsilon_0^{A}-h_A(\xi)\right)\Phi_{0}=0,\quad E_{0}=E-\epsilon_0^{A}-\epsilon_0^{a},
\end{align}
while the many-body Green's function is obtained from $H_{B}$ in (\ref{eq500}) assuming $\Delta V_{B}=0$,
\begin{align}\label{eq405}
\mathbf G_{I}(\mathbf r_{xA},\mathbf \xi,\mathbf r_{bB},E)=\left(E-T_{x}-T_{b}-h_{A}(\xi)-V_{xA}(\mathbf r_{xA},\xi)-U_{bB}(\mathbf r_{bB})\right)^{-1}.
\end{align}
We use the subscript $_I$ to indicate that the Green's function above corresponds to the \emph{indirect} measurement, in order to distinguish it from the \emph{direct} Green's function (\ref{eq401}), which does not depend on $\mathbf r_{bB}$. The above expression exemplifies what can be considered the essence of the spectator approximation:  the factorization of the Green's function in a product of operators in the $b$ and $x$-$A$ spaces. More explicitly, (\ref{eq405}) can be written as
\begin{align}\label{eq330}
 \nonumber \mathbf G_{I}(E)&=\int d^3 k_{b}\ket{\chi_{b}(\mathbf r_{bB},\mathbf k_{b})}\bra{\chi_{b}(\mathbf r_{bB},\mathbf k_{b})}\otimes \sum_{ij}\ket{\Phi_{i}(\xi)}\bra{\Phi_{j}(\xi)}\\
 \nonumber &\times\braket{\Phi_{i}(\xi)|\left(E_{i}-E_{b}-T_{x}(\mathbf{r}_{xA})-V_{xA}(\mathbf{r}_{xA},\xi)\right)^{-1}|\Phi_{j}(\xi)}\\
  \nonumber &=\int d^3 k_{b}\ket{\chi_{b}(\mathbf r_{bB},\mathbf k_{b})}\bra{\chi_{b}(\mathbf r_{bB},\mathbf k_{b})}\\
  &\otimes \sum_{ij}\ket{\Phi_{i}(\xi)}\left(\left(E_{i}-E_{b}-T_{x}(\mathbf{r}_{xA}))\right)\delta_{ij}-V_{ij}(\mathbf{r}_{xA})\right)^{-1}\bra{\Phi_{j}(\xi)}
\end{align}
where $E_{i}=E-\epsilon_{i}^{A}$, and we have used the fact that the Green's function is diagonal in the basis of $b$ states $\chi_{b}$ defined by
\begin{align}
\big(E_{b}-T_{b}-U_{bB}(\mathbf r_{bB})\big)\,\chi_{b}(\mathbf r_{bB},\mathbf k_{b})=0;\quad \left(E_{b}=\frac{\hbar^{2}k_{b}^{2}}{2\mu_{b}}\right).
\end{align}
The matrix elements $V_{ij}$ of the many-body potential were defined in Eq. (\ref{eq306}).
Comparing with (\ref{eq401}), we obtain
\begin{align}
\mathbf G_{I}(\mathbf r_{xA},\xi,\mathbf r_{bB},E)=\int d^3 k_{b}\mathbf G(\mathbf r_{xA},\xi,E-E_{b})\,\mathcal P_{b}(\mathbf r_{bB},\mathbf k_{b}),
\end{align}
where we have introduced the projection operator
\begin{align}
\mathcal P_{b}(\mathbf k_{b})=\ket{\chi_{b}(\mathbf r_{bB},\mathbf k_{b})}\bra{\chi_{b}(\mathbf r_{bB},\mathbf k_{b})}.
\end{align}
Eq. (\ref{eq408}) can now be rewritten,
\begin{align}\label{eq557}
\Psi=\Psi_{0}+\int d^3 k_{b}\,\mathbf G(E-E_{b})\,\mathcal P_{b}(\mathbf k_{b})\left[V_{xA}(\mathbf r_{xA},\xi)+U_{bA}(\mathbf r_{bA})\right]\Psi_{0}=\Psi_{0}+\Psi_{xA},
\end{align}
where the factorized Green's function is consistent with a process in which two systems (the $x$-$A$ system on the one hand and the $b$ fragment on the other hand) scatter independently. 
Let us now obtain the one-body elastic scattering (defined as the process in which the nucleus $A$ remains in its ground state) wave function corresponding to an experimental situation in which the fragment $b$ has been observed with momentum $\mathbf k_{b}$. This is obtained by projecting the many-body wavefunction (\ref{eq557}) onto the corresponding observed state $\ket{\Phi_{0}\,\chi_{b}(\mathbf k_{b})}$,
\begin{align}\label{eq409}
\nonumber &\int \Phi_{0}(\xi)\,\chi_{b}(\mathbf r_{bB},\mathbf k_{b})\Psi(\mathbf r_{xA},\mathbf r_{bB},\xi)\,d^3r_{bB}\,d\xi\equiv \bra{\Phi_{0}\,\chi_{b}(\mathbf k_{b})}\Psi\\
\nonumber &=\bra{\Phi_{0}\,\chi_{b}(\mathbf k_{b})}\Psi_{0}+\bra{\Phi_{0}\,\chi_{b}(\mathbf k_{b})}\mathbf G_{I}(E)\left[V_{xA}(\mathbf r_{xA},\xi)+U_{bA}(\mathbf r_{bA})\right]\Psi_{0}\\
 \nonumber &=\bra{\chi_{b}}\phi_{a}F+\bra{\Phi_{0}}\mathbf G(E-E_{b})\bra{\chi_{b}}\left[V_{xA}(\mathbf r_{xA},\xi)+U_{bA}(\mathbf r_{bA})\right]\Psi_{0}\\
 &=\bra{\chi_{b}}\phi_{a}F+G(E_{0}-E_{b})\bra{\chi_{b}}\left[\mathcal V(E-E_{b})+U_{bA}(\mathbf r_{bA})\right]\phi_{a}F,
\end{align}
where we have used (\ref{eq406}). We can now write the \emph{indirect} elastic channel wavefunction for the fragment $x$ \emph{in terms of single-particle operators},
\begin{align}\label{eq325}
\nonumber \psi_{0}^{I}(\mathbf r_{xA})&=\psi^{HM}+G(E_{0}-E_{b})\left[\mathcal V(E-E_{b})\psi^{HM}+\bra{\chi_{b}} U_{bA}(\mathbf r_{bA})\,\phi_{a}F\right]\\
\nonumber &=\psi^{HM}(\mathbf r_{xA})+G(\mathbf r_{xA},E_{0}-E_{b})\\
&\times \left[\mathcal V(\mathbf r_{xA},E-E_{b})\psi^{HM}(\mathbf r_{xA})+\int \chi^{*}_{b}(\mathbf r_{bB},\mathbf k_{b})\,U_{bA}(\mathbf r_{bA})\,\phi_{a}(\mathbf r_{xb})F(\mathbf r_{aA})\,d r^3_{xb} \right],
\end{align}
or, equivalently,
\begin{align}\label{eq420}
\nonumber \psi_{0}^{I}(\mathbf r_{xA})&=\psi^{HM}+G(E_{0}-E_{b})\bra{\chi_{b}}\big(\,\mathcal V(E-E_{b})+ U_{bA}(\mathbf r_{bA})\,\big)\,\phi_{a}F\\
&=\psi^{HM}(\mathbf r_{xA})+G(\mathbf r_{xA},E_{0}-E_{b})\int \chi^{*}_{b}(\mathbf r_{bB},\mathbf k_{b})\left(\mathcal V(\mathbf r_{xA},E-E_{b})+U_{bA}(\mathbf r_{bA})\right)\,\phi_{a}(\mathbf r_{xb})F(\mathbf r_{aA})\,d r^3_{xb} ,
\end{align}
where the single-particle Green's function is the one defined in Eq. (\ref{eq413}), and we have introduced the Hussein-McVoy wavefunction \cite{husseinInclusiveProjectileFragmentation1985},
\begin{align}\label{eq:HM}
\psi^{HM}(\mathbf r_{xA})=\int \chi^{*}_{b}(\mathbf r_{bB},\mathbf k_{b})\phi_{a}(\mathbf r_{xb})F(\mathbf r_{aA})\,d r^3_{xb}.
\end{align}
We can now use Eq. (\ref{eq407}) to obtain the total indirect reaction cross section of the fragment $x$ with the nucleus $A$,
\begin{align}\label{eq412}
\frac{d\sigma^{I}_{R}(E,E_{b})}{dE_b}=\frac{2\mu}{\hbar^2 k_{x}}\braket{\psi^{I}_{0}|\textrm{Im}\mathcal V(E-E_{b})
|\psi^{I}_{0}}\,\rho(E_b),
\end{align}
where $\rho(E_b)$ is the density of states of the fragment $b$ (in units $[\rho]=\text{MeV}^{-1}\text{fm}^{-3}$), and $\mu$ is the reduced mass of the $x$-$A$ system. The above expression is the main result of this section, and it can be used to compute the cross section for the inclusive reaction $a+A\to b+(x+A)$, where only the fragment $b$ is detected.
This result, identical to the one obtained in \cite{potelEstablishingTheoryDeuteroninduced2015} with a quite different method, is the essence of the Green's Function Transfer (GFT) formalism. 

In practical applications, the optical potential is either obtained from a phenomenological fit (\cite{potelEstablishingTheoryDeuteroninduced2015,a.ratkiewiczNeutronCaptureExotic2019}), or calculated within some structure formalism (Sects. \ref{SVA}, \ref{SVB}, see also \cite{rotureauMergingInitioTheory2020,barrancoMathrmLiReactionSpecific2020}). In addition, a distorted wave associated with an auxiliary optical potential is used instead of the free plane wave $F(\mathbf r_{aA})$, with the  corresponding modifications indicated in (\ref{eq321}). Then, a partial wave expansion of the wavefunction and the potential allows for the calculation of the reaction (absorption) cross section as a function of the angular momentum of the composite $x$-$A$ system. 
 
The expression (\ref{eq412}) corresponds to the calculation of the reaction cross section, i.e., of the  the \emph{total} flux removed from the elastic channel of the $x$-$A$ system. As we discussed in Sect. \ref{SOP}, this flux is feeding \emph{all} energy-conserving excited states of the $x$-$A$ system. Within this context, this calculation is said to be \emph{inclusive} with respect to the states of the $x$-$A$ system, as opposed to the \emph{exclusive} calculations where a specific final state of the $x$-$A$ system is selected (see Sect. \ref{SVI}).  However, it is often the case that only one excited state exists at a particular excitation energy, and the inclusive cross section is exhausted by the population of this single state. This is always the case for bound states (see Sect. \ref{SVB}), and can also approximately happen when a resonance is rather well isolated (see Sect. \ref{SVA}). 

In the spirit of the use of indirect reactions invoked in these lectures, which is to benefit from the (indirect) cross section (\ref{eq412}) to extract information about the $x$-$A$ system, an interesting feature of the GFT formalism is that the cross section is explicitly computed in terms of the self energy $\mathcal V$ which determines the structure of the $x$-$A$ system. Therefore, a theoretical calculation of the self energy of the $x-A$ system can be used to predict the cross section (\ref{eq412}), and  be directly compared with experimental data in order to assess the validity of the structure formalism. The full consistency of the structure calculation (embodied in the self energy/optical potential $\mathcal V$) with the reaction calculation (the Green's function/propagator $G$) is enforced by  Eq. (\ref{eq413}). In Sects \ref{SVA} and \ref{SVB} we present examples of this procedure.
\subsubsection{Connection with the DWBA}\label{CDWBA}
As the energy $E_b$ of the fragment $b$ becomes larger, the argument of the optical potential in Eq. (\ref{eq412}) will eventually be negative, and the corresponding cross section will be associated to the population of bound states of the $x+A$ system, i.e.,  below the threshold for emitting the fragment $x$. At negative energies, the optical potential becomes real (see \cite{dickhoffRecentDevelopmentsOptical2019}), which we can express by writing 
\begin{align}\label{eq419}
\mathcal V(E)=\lim_{\epsilon\to 0}\text{Re}\mathcal V(E)+i\epsilon,\quad \text{for }E<0.  
\end{align}
For a discrete (bound) spectrum, the spectral decomposition of the Green's function (Lehmann representation, see e.g. \cite{dickhoffManyBodyTheory2005}), is then
\begin{align}\label{eq421}
G(E)=\lim_{\epsilon\to 0}\sum_i\frac{\ket{\psi_i}\bra{\psi_i}}{E-E_i+i\epsilon},
\end{align}
where $\psi_i,E_i$ are the eigenstates and eigenenergies of the $x-A$ system. The coordinate representation of the above Green's function is obtained making use of the wavefunctions of the eigenstates\footnote{Note that, with this definition, the Green's function has units of $[G]=\text{MeV}^{-1}\text{fm}^{-3}$. In the context of the convention for non local operators stated in the introduction, it thus corresponds to the primed operator  in the definition (\ref{eq:321}). In other words, the application of this Green's function to an arbitrary function of the coordinates has to be interpreted as
\begin{align*}
G(E,\mathbf r_{xA},\mathbf r'_{xA})f(\mathbf r_{xA})=\int d^3 r'_{xA}\,G(E,\mathbf r_{xA},\mathbf r'_{xA})f(\mathbf r'_{xA}),
\end{align*}
at variance with the convention used in these lectures.
}
\begin{align}\label{eq434}
G(E,\mathbf r_{xA},\mathbf r'_{xA})=\lim_{\epsilon\to 0}\sum_i\frac{\psi_i(\mathbf r_{xA})\psi_i^*(\mathbf r'_{xA})}{E-E_i+i\epsilon}.
\end{align}

\begin{figure}[h]
	\centerline{\includegraphics[width=13cm]{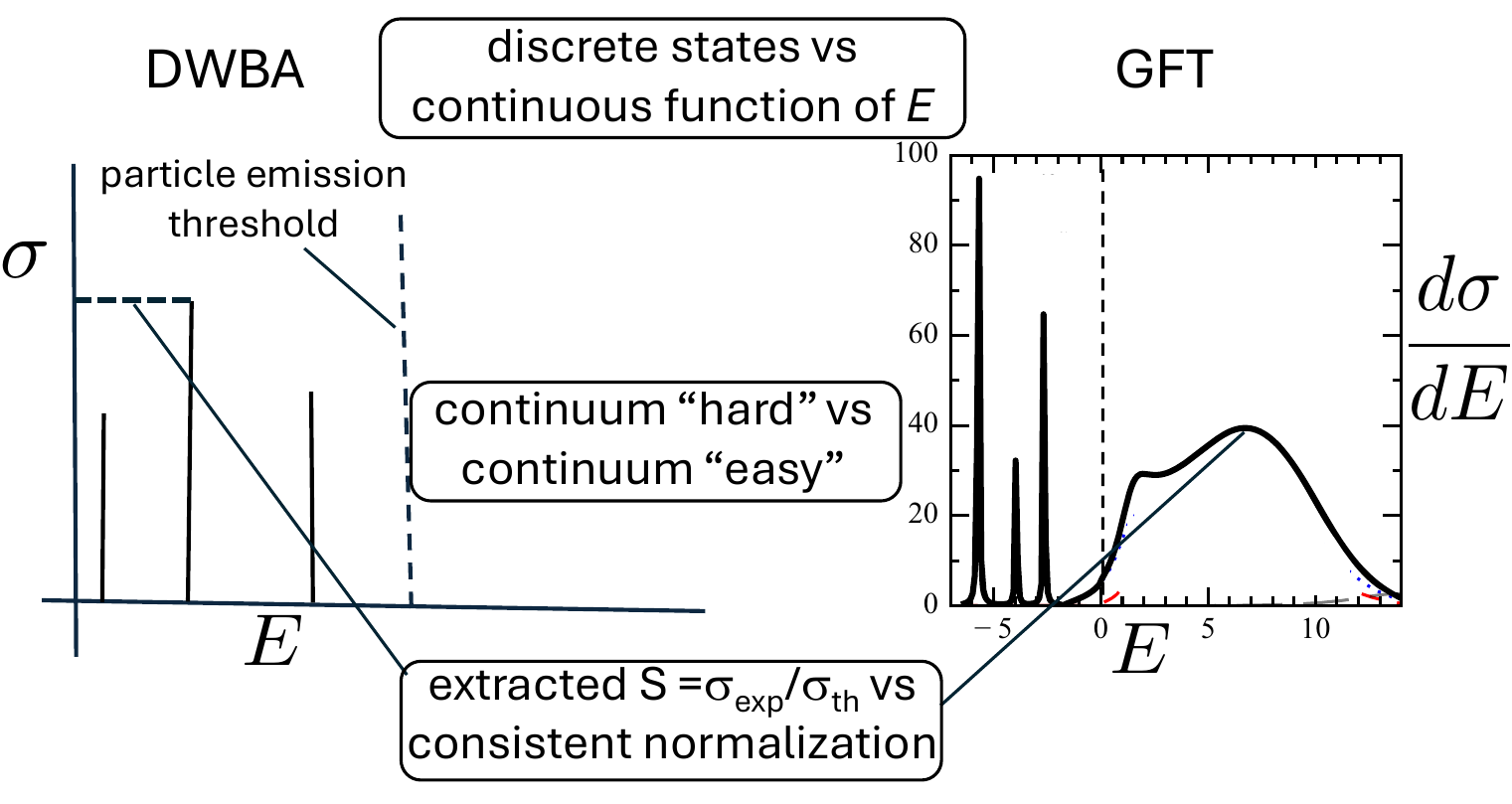}}
	\caption{Schematic depiction of the distinctive features of the Distorted Wave Born approximation (DWBA) as compared with the Green's Function Transfer (GFT) formalism.}
	\label{fig2} 
\end{figure}

Let us further note that, if the optical potential $\mathcal V$ is energy-dependent (i.e., whenever the interaction between $x$ and $A$ couples the ground state of $A$ with its excited states, see Sect. \ref{SOP}), the eigenfunctions in (\ref{eq421}) are not normalized to 1, but they rather verify (\cite{dickhoffManyBodyTheory2005})
\begin{align}\label{eq422}
\braket{\psi_i|\psi_i}=\left(1-\frac{\partial \mathcal V}{\partial E}\right)^{-1}=S_i,
\end{align}
where $S_i$ is the spectroscopic amplitude of the state $i$. This is an important point: the single-particle Green's function contains the information about the proper normalization of the single-particle states (the residues of the poles of the Green's function). Since the nucleus is a correlated many-body system, as testified by the energy dependence of the optical potential, the single-particle states are not orthonormal, and the normalization of the wavefunctions is not trivial. Because the GFT makes use of a Green's function consistent with the optical potential (see Eq. (\ref{eq413})), the resulting cross sections are properly normalized, and the corresponding absolute values can be directly compared with experimental data (see Fig. \ref{fig2}).  

Let us now write down the different terms arising in the evaluation of the cross section (\ref{eq412}) using (\ref{eq420}),
\begin{align}\label{eq423}
\frac{d\sigma^{I}_{R}(E,E_{b})}{dE_b}=\frac{2\mu}{\hbar^2 k_{x}}(A_1+A_2+A_3)\,\rho(E_b),
\end{align}
where
\begin{align}\label{eq424}
A_1=\braket{\psi^{HM}|\text{Im}\mathcal V|\psi^{HM}}=\lim_{\epsilon \to 0}\epsilon\,\braket{\psi^{HM}|\psi^{HM}}=0,
\end{align}
\begin{align}\label{eq425}
A_2=\braket{\psi^{xA}|\text{Im}\mathcal V|\psi^{xA}},
\end{align}
with
\begin{align}\label{eq426}
\psi^{xA}=G(E_{0}-E_{b})\bra{\chi_{b}}\left(\mathcal V(E-E_{b})+ U_{bA}(\mathbf r_{bA})\right)\,\phi_{a}F,
\end{align}
and the cross term is
\begin{align}\label{eq427}
A_3=2\text{Re}\braket{\psi_{xA}|\text{Im}\mathcal V|\psi^{HM}}.
\end{align}
We now express the term $A_2$ as
\begin{align}\label{eq428}
\nonumber A_2&=\lim_{\epsilon\to 0}\epsilon\left[G(E_{0}-E_{b})\bra{\chi_{b}}\left(\mathcal V(E-E_{b})+ U_{bA}(\mathbf r_{bA})\right)\,\phi_{a}F\right]^{\dagger}\,\\
\nonumber &\times G(E_{0}-E_{b})\bra{\chi_{b}}\left(\mathcal V(E-E_{b})+ U_{bA}(\mathbf r_{bA})\right)\,\phi_{a}F\\
&=\lim_{\epsilon\to 0} \sum_i \frac{\epsilon}{(E_0-E_b-E_i)^2+\epsilon^2}|T_i|^2=\sum_i|T_i|^2\delta(E_0-E_b-E_i)
\end{align}
where we have used (\ref{eq421}), and we have defined the transfer $T$-matrix to the bound state $\psi_i$,
\begin{align}\label{eq429}
T_i=\braket{\psi_i\chi_b|\,\mathcal V(E-E_{b})+ U_{bA}(\mathbf r_{bA})\,|\phi_a F}.
\end{align}
The  cross term $A_3$ also vanishes, since it is proportional to the energy-dependent term
\begin{align}\label{eq430}
\text{Re}\left(\lim_{\epsilon\to 0}\sum_i \frac{\epsilon}{(E_0-E_b-E_i)+i\epsilon}\right)=0.
\end{align}
The cross section is then 
\begin{align}\label{eq431}
\frac{d\sigma^{I}_{R}(E,E_{b})}{dE_b}=\frac{2\mu}{\hbar^2 k_{x}}\sum_i|T_i|^2\delta(E_0-E_b-E_i)\,\rho(E_b),
\end{align}
and has the same structure as the transfer cross section to bound states computed in DWBA (see Sect. \ref{DWBA}, see also \cite{potelEstablishingTheoryDeuteroninduced2015}).

 In practice, when making a DWBA calculation the wavefunction $\psi_i$ in Eq. (\ref{eq429}) is usually taken to be a solution \emph{normalized to 1} of the Schr\"odinger equation for the bound state $i$ in the potential $\mathcal V(E-E_{b})$, which is approximated by a simple central potential, often a Woods-Saxon with a reasonable radius and a depth fitted to reproduce the binding energy of the state $i$. The explicit connection of the resulting cross section with the correlated many-body system, enforced by Eq. (\ref{eq413}), is then lost. In particular, the value of the single-particle strength has to be extracted by comparing the calculated cross section with the experimental one, (see Fig. \ref{fig2}) a procedure which is rendered ambiguous by the arbitrariness in the choice of the wavefunction $\psi_i$ to be used in Eq. (\ref{eq429}). 
 
 Let us stress that, although common practice, the procedure that has been just described is not an inherent part of the DWBA. It is always possible to use in the $T$-matrix a wavefunction consistent (both in terms of normalization and of spatial dependence) with a microscopic self energy $\mathcal V$ calculated within some structure formalism of choice, to be also used for the calculation of the $T$-matrix (\ref{eq429}). The resulting cross section would then be equivalent to the one computed with the GFT formalism.

Let us finally comment on the population of the continuum. The spectral decomposition of the Green's function associated with the continuum part of the spectrum of the $x-A$ system is
\begin{align}\label{eq433}
G(E,\mathbf r_{xA},\mathbf r'_{xA})=\lim_{\epsilon\to 0}\int dE_c\,\rho(E_c)\,\frac{\chi(\mathbf r_{xA},E_c)\chi^*(\mathbf r'_{xA},E_c)}{E-E_c+i\epsilon},
\end{align}
where $\chi$ are continuum wavefunctions, and $\rho(E_c)$ is the density of states of the continuum. This expression should  substitute Eq. (\ref{eq434}) in the expression for the Green's function to be used in Eq. (\ref{eq428}). The resulting expression would then not have the form of a sum over DWBA $T$-matrices corresponding to the population of a discrete set of final bound states, but would rather involve an integral over the continuum states.  The essentially discrete nature of the DWBA renders somehow unnatural the description of the population of continuum states,  although several methods exist to discretize the continuum and describe the associated cross sections \cite{Austern1987CDCC,Kamimura1986PTPS,Yahiro2012PTEP} (see Fig. \ref{fig2}). Let us also note that the proper normalization of the continuum wavefunctions is not trivial, while the GFT formalism takes naturally care of that. In addition, since the optical potential $\mathcal V(E>0)$ has a finite imaginary part, the terms $A_1,A_3$ do not vanish, and the exact connection with the DWBA is lost. Finally, let us point out that the DWBA cannot properly account for the coherent contribution to the cross section of overlapping resonances of the same spin and parity, while this possibility is properly embodied in the Green's function used in the GFT.  

\section{Applications}
As examples of the GFT formalism, we will discuss two applications, associated with very different systems. In both cases, the implementation of the GFT implies the calculation of the self energy $\mathcal V$ within some structure formalism of choice, and the obtention of the associated Green's function by inverting the Hamiltonian matrix according to Eq. (\ref{eq413}). The consistency between structure ($\mathcal V$) and reactions $(G)$, is thus enforced. From a technical point of view, this implies the development of methods and tools that allow us  to deal with the non quite standard situation of having a non-local potential $\mathcal V$, resulting from most self energy calculations. These methods have been developed, allowing for the computation of the solution of the Schr\"odinger equation for non local potentials, and the obtention of Green's functions from matrix inversion, for positive and negative energies, as well as for charged and neutral particles (see, e.g., \cite{sargsyanMicroscopicOpticalPotentials2024}). 

The first example will describe the population of the low-lying spectrum of the unbound system $^{10}$Li in the $^{9}$Li$(d,p)$ reaction. This process serves as a good illustration of the description of the population of the continuum in indirect reactions within the GFT formalism. 

The second example will be the population of the $^{41}$Ca nucleus in the $^{40}$Ca$(d,p)$ reaction, which is a good example of the use of the GFT formalism to describe the population of bound states in a stable nucleus.

\subsection{\texorpdfstring{$^{9}$Li$(d,p)^{10}$Li}{x} with NFT}\label{SVA}

The unbound nucleus $^{10}$Li sits at the crossroads of shell evolution and halo structure. 
In the $N{=}7$ chain, the normal ordering places the $1p_{1/2}$ below the $2s_{1/2}$ orbit, as seen in $^{12}$B and $^{13}$C; yet the paradigm of parity inversion in $^{11}$Be --- where the $1/2^+$ bound state lies below the $1/2^-$ --- suggests that the isotone $^{10}$Li should display a near--threshold \emph{virtual} $1/2^+$ configuration in competition with a low--lying $1/2^-$ \emph{resonance}. When coupled to the odd $3/2^-$ proton in  $^9$Li, these states give rise to a $1^+$ and $2^+$ doublet associated with the  $1/2^-$ resonance and a $1^-$, $2^-$ doublet for the  $1/2^+$ state. 
Establishing this has direct consequences for the large $s$--wave component of the $^{11}$Li halo and for a unified, many--body description of this region. 
The reaction $^{9}$Li$(d,p)^{10}$Li is a specific probe of single--neutron motion built on the $^{9}$Li core: by measuring proton spectra and angular distributions, one gains access to the energy dependence and partial--wave content of the $n{+}{}^{9}$Li continuum. Within this context, and in the spirit of these lectures, the $^{9}$Li$(d,p)^{10}$Li reaction can be viewed as an \emph{indirect} measurement of the $n{+}{}^{9}$Li system, the \emph{direct} measurement of neutron scattering on $^{9}$Li being experimentally essentially unfeasible due to the short life of $^{9}$Li (178 ms) and of the neutron (614 s). 

In the NFT approach, the structure calculations in \cite{barrancoMathrmLiReactionSpecific2020} dress the neutron--core motion with strong particle--vibration coupling to the quadrupole phonon of the $^{9}$Li core ($\beta_2=0.72$, $\hbar\omega_2 \simeq 3.4$~MeV), including relevant two--phonon (anharmonic) contributions in the $d_{5/2}$ channel.
The output is an energy--dependent, nonlocal self--energy $\Sigma(E)$ which shifts, fragments, and reshapes the single--particle configurations. The resulting  dressed continuum states $\widetilde{j}^\pi$ exhibit renormalized energies, widths, spectroscopic amplitudes, and --- crucially for transfer --- the proper \emph{radial form factors}. 
This same $\Sigma(E)$ is used to provide the single--particle Green's function $G(E)=(E-T-\Sigma)^{-1}$ used in the GFT approach  (Sec.~\ref{Sec2}), ensuring that \emph{structure and reaction consistent and treated on the same footing}, allowing predictions of \emph{absolute} transfer cross sections without ad hoc normalizations.

An $s$--wave \emph{virtual} state is encoded in a large, negative scattering length $a$; equivalently the $s$--wave phase shift behaves as
\begin{equation}
\delta_0(E) \xrightarrow[E\to 0]{} -k a, \qquad E=\frac{\hbar^2 k^2}{2\mu},
\end{equation}
implying an energy scale $\varepsilon \sim \hbar^2/(2\mu a^2)$ that controls the near--threshold enhancement. 
For a narrow resonance at $E_r$ one has the familiar relation
\begin{equation}
\Gamma_{j^\pi}(E_r)= 2\left(\frac{d\delta_{j^\pi}}{dE}\bigg|_{E_r}\right)^{-1}\!,
\end{equation}
which we will use below to interpret the $p$-- and $d$--wave structures.

\begin{figure}[t]
  \centering
  \begin{minipage}[t]{0.48\linewidth}
    \includegraphics[width=\linewidth]{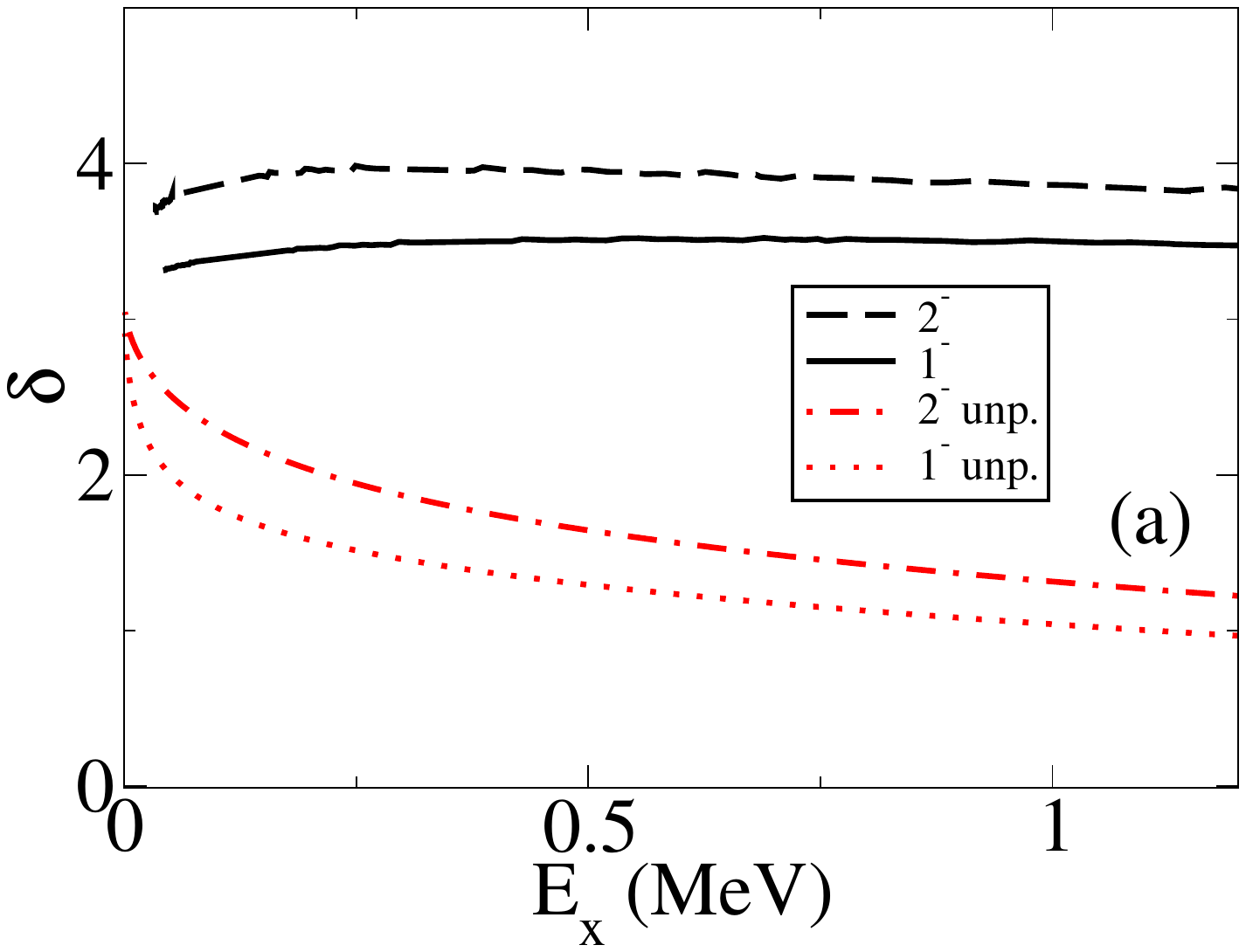}
  \end{minipage}\hfill
  \begin{minipage}[t]{0.48\linewidth}
    \includegraphics[width=\linewidth]{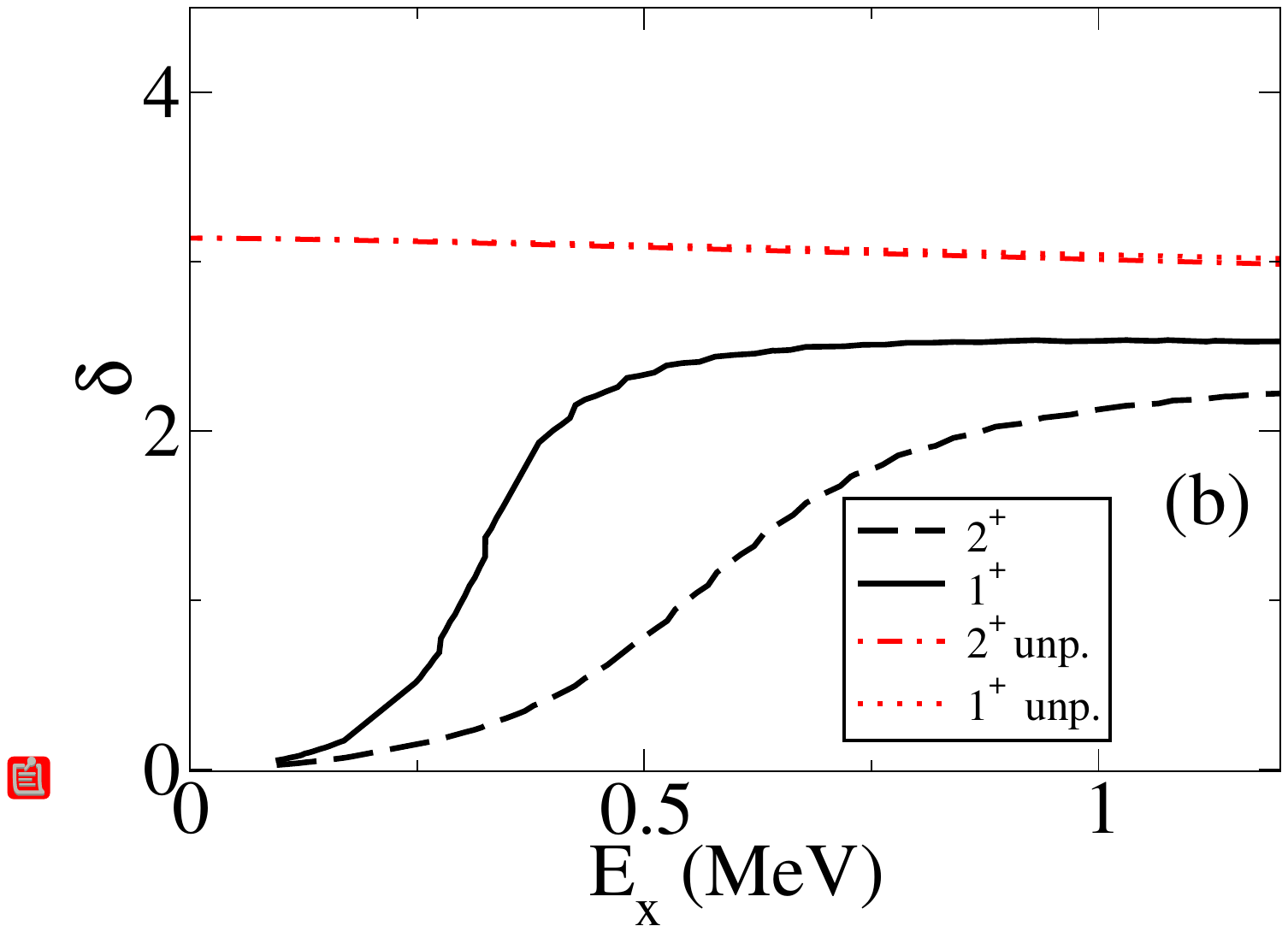}
  \end{minipage}\\[0.8ex]
  \begin{minipage}[t]{0.48\linewidth}
    \includegraphics[width=\linewidth]{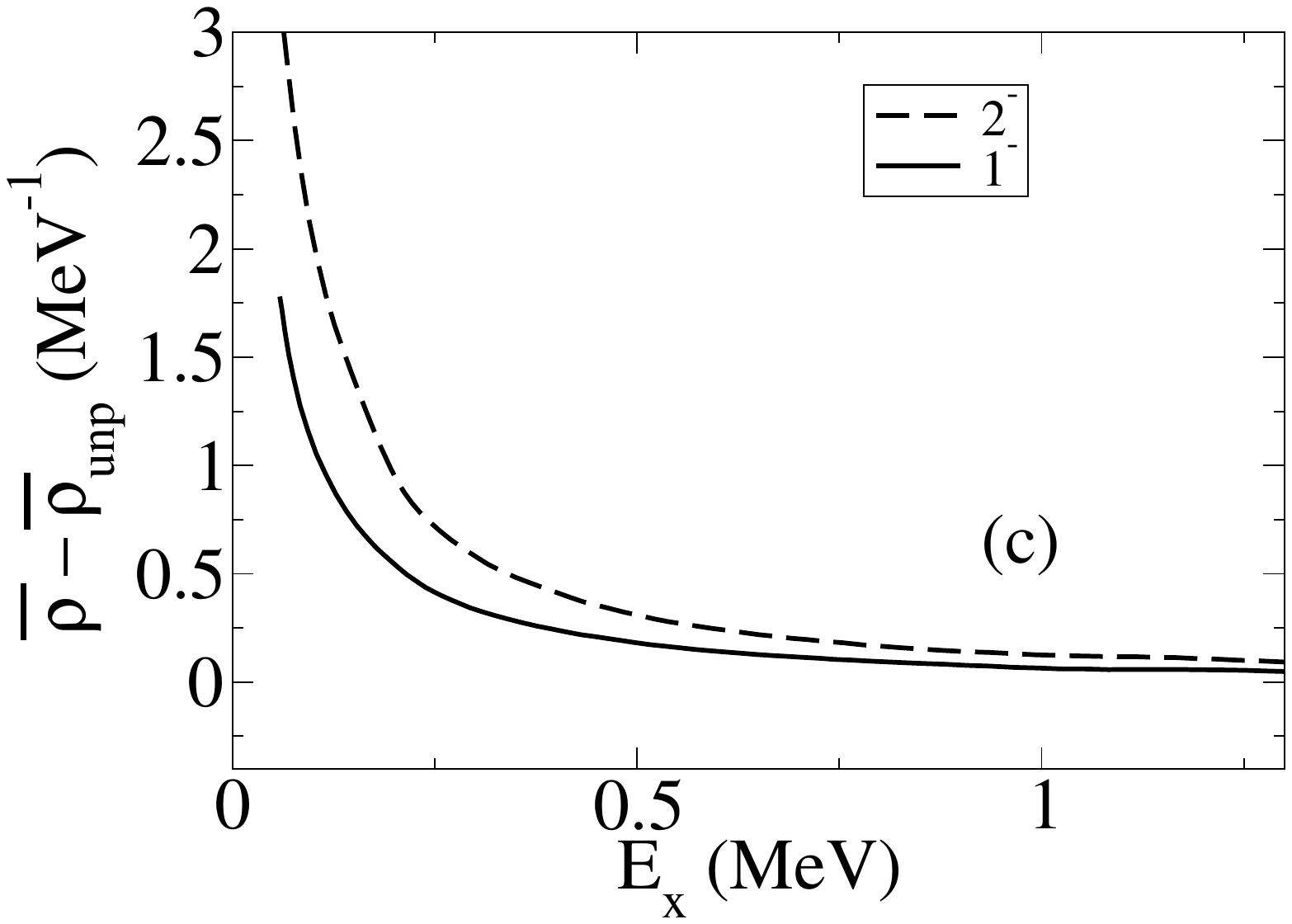}
  \end{minipage}\hfill
  \begin{minipage}[t]{0.48\linewidth}
    \includegraphics[width=\linewidth]{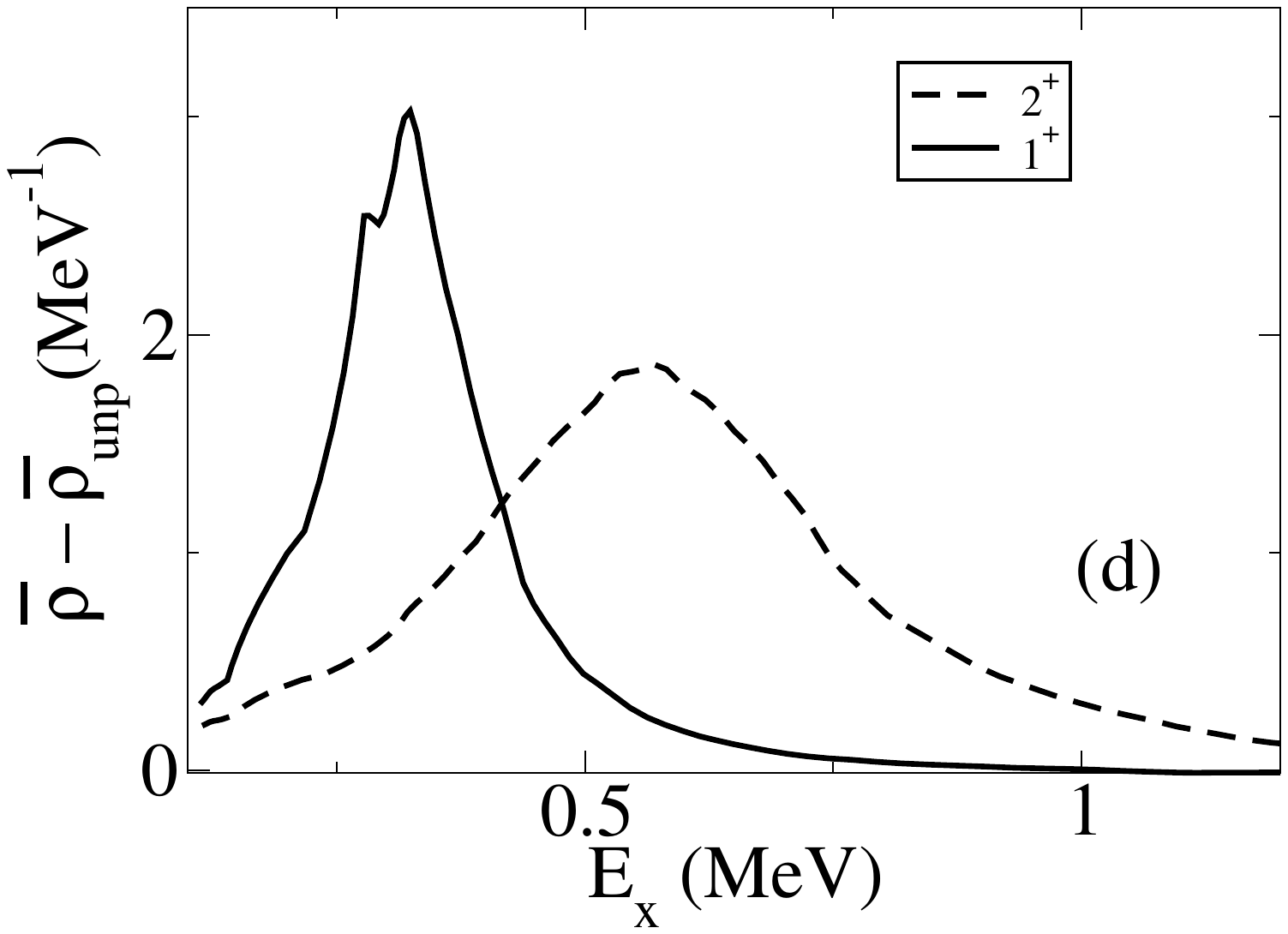}
  \end{minipage}
  \caption{Phase shifts (top) and spectral functions (bottom) for $^{10}$Li from the renormalized NFT calculation of \cite{barrancoMathrmLiReactionSpecific2020}, compared with the unperturbed (``unp.'') case. 
  (a) Negative parity: the $2^-$ and $1^-$ channels develop a low--lying $\widetilde{1/2}^-$ resonance at $\sim$0.5~MeV. 
  (b) Positive parity: the $s$--wave becomes virtual (large negative $a$) and a broad $\widetilde{5/2}^+$ resonance appears at a few MeV. 
  (c,d) Differences of spectral functions highlight the redistribution of strength due to particle--vibration coupling.}
  \label{fig:prc2}
\end{figure}

Figures~\ref{fig:prc2} (a,b) show the calculated phase shifts for the negative-- and positive--parity channels, compared to the bare mean--field (``unp.'') behaviour. 
The dressing reverses the slope of the $2^-$ and $1^-$ phase shifts near threshold, signalling the formation of a low--lying $\widetilde{1/2}^-$ resonance around $0.45$--$0.50$~MeV with a width of about $0.3$--$0.35$~MeV. 
In the positive--parity sector, the $s_{1/2}$ strength is pushed into the continuum and acquires a \emph{virtual} character, with a scattering length $a \simeq -8$~fm (corresponding scale $\varepsilon \approx 0.3$~MeV), while the $d_{5/2}$ channel develops a broad $\widetilde{5/2}^+$ resonance at a few MeV due to strong coupling to the quadrupole phonon and two--phonon states. 
Panels (c,d) present the corresponding \emph{spectral functions}, i.e.\ the differences between the renormalized and bare level densities, $ \bar\rho_{j^\pi}(\omega) - \bar\rho^{\rm unp}_{j^\pi}(\omega) \propto d\delta_{j^\pi}/d\omega$, which make the redistribution of strength transparent.

\begin{figure}[t]
  \centering
  \includegraphics[width=.9\linewidth]{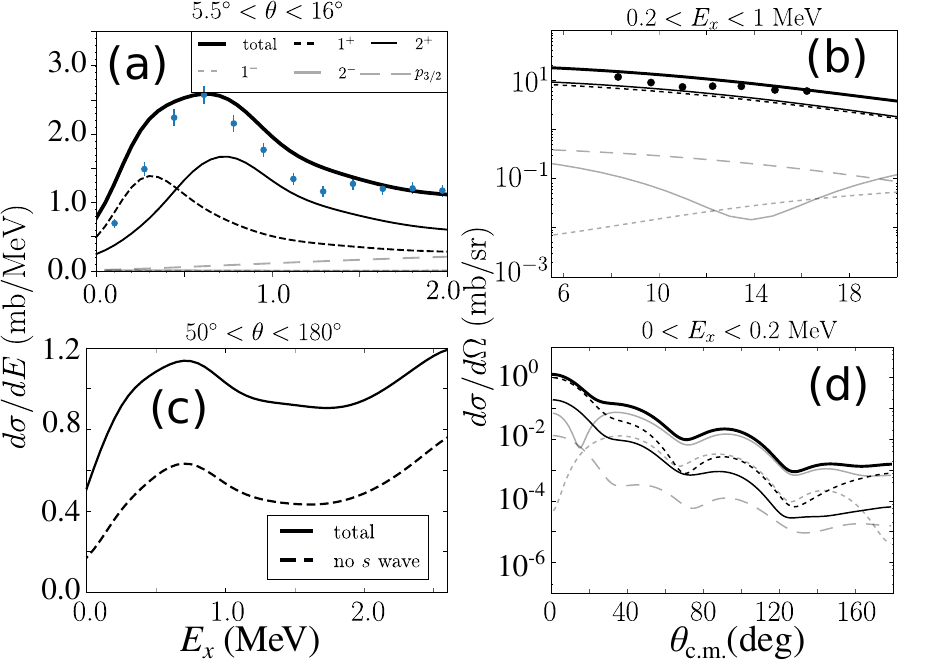}
  \caption{Absolute $(d,p)$ predictions from \cite{barrancoMathrmLiReactionSpecific2020}: (a) strength function folded over the \emph{forward} window $5.5^\circ$--$16.5^\circ$ (dominated by $p$--wave); (b) corresponding angular distributions in a representative energy interval; (c) strength integrated over \emph{backward} angles $50^\circ$--$180^\circ$, where the $s$--wave becomes prominent; (d) low--energy angular distribution ($E{\lesssim}0.2$~MeV) emphasizing the virtual $s$--wave at $\theta_{\rm c.m.}\gtrsim 40^\circ$.}
  \label{fig:prc3}
\end{figure}

As stated above, these predictions cannot be directly addressed  with a neutron scattering experiment.  Instead, one can resort to the indirect $(d,p)$ measurement in inverse kinematics, making use of a radioactive $^9$Li beam on a deuterated target. The calculated self energy $\Sigma(E)$  at the basis of these results is implemented in the expressions (\ref{eq420}) and (\ref{eq412}) --where it takes the place of the optical potential $\mathcal V(E)$-- to compute  the absolute single-- and double--differential cross sections and fold it over the experimental resolutions. 
The results are shown in Fig.~\ref{fig:prc3}: when the yield is integrated over the forward angular window $5.5^\circ \le \theta_{\rm c.m.} \le 16.5^\circ$ --- the acceptance of the high--statistics measurement reanalysed in \cite{barrancoMathrmLiReactionSpecific2020} --- the spectrum is dominated by the $p$--wave resonance and exhibits an \emph{apparent} absence of $s$--wave strength near threshold. 
However, if one inspects the angular distributions for $\theta_{\rm c.m.}\gtrsim 40^\circ$, or integrates over a backward window (e.g.\ $50^\circ$--$180^\circ$), the virtual $s$--wave contribution emerges clearly, with a magnitude comparable to the $p$--wave component close to threshold. 
Thus the seemingly contradictory inferences regarding parity inversion are reconciled once momentum matching and acceptance effects are accounted for within a consistent structure--reaction framework.

The dressed $\widetilde{1/2}^+$ state in $^{10}$Li is predominantly $s_{1/2}$ with a small admixture of $(d_{5/2}\!\otimes\!2^+)$; the $\widetilde{1/2}^-$ resonance contains $(p_{1/2}\otimes2^+)$ components; and the $\widetilde{5/2}^+$ resonance around $3$--$4.5$~MeV is strongly many--body in nature due to two--phonon couplings. 
This same mechanism unifies the parity inversion in $^{11}$Be, the large $s$--wave content of the $^{11}$Li halo, and the normal ordering in $^{12}$B/$^{13}$C. 
From the practical viewpoint of transfer reactions, the decisive point is that \emph{the same self--energy} that fixes the spectroscopy also fixes the \emph{radial form factors} entering the cross sections, removing the traditional ambiguity of separate DWBA fits.

\begin{figure}[t]
  \centering
  \includegraphics[width=.8\linewidth]{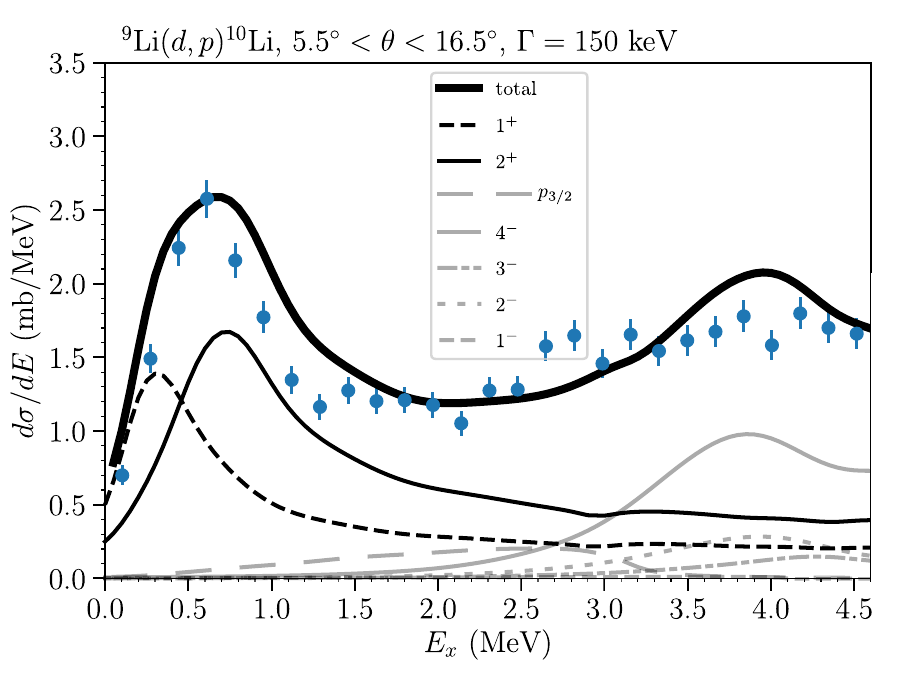}
  \caption{Results obtained from the analysis of the $^9$Li($d,p$)$^{10}$Li reaction  within the renormalized NFT\,+\,GFT framework: (i) parity inversion encoded in a virtual $\widetilde{1/2}^+$ (large negative scattering length) alongside a low--lying $\widetilde{1/2}^-$ resonance; (ii) dressing by the strong $2^+$ vibration of the $^{9}$Li core and associated two--phonon effects that redistribute $d_{5/2}$ strength; (iii) momentum matching in $(d,p)$ that suppresses the $s$--wave at forward angles yet enhances it at $\theta_{\rm c.m.}\gtrsim 40^\circ$.}
  \label{fig:Li10_schematic}
\end{figure}

We retain Fig.~\ref{fig:Li10_schematic} as a compact account of the story: the near--threshold virtual $\widetilde{1/2}^+$ and the low--lying resonant $\widetilde{1/2}^-$ shaped by coupling to the strong $2^+$ core vibration, and the momentum--matching selectivity that hides (forward angles) or reveals (backward angles) the $s$--wave contribution in $(d,p)$.

\begin{figure}  
  \centering
  \includegraphics[width=.75\linewidth]{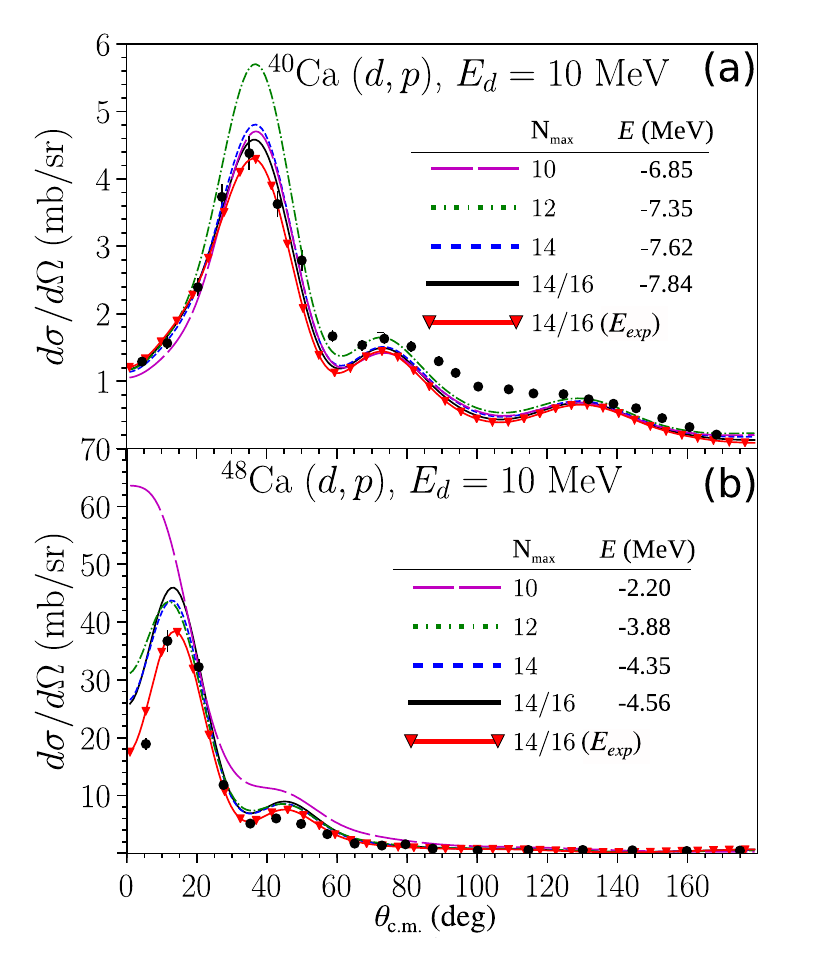}
  \caption{Angular differential cross sections for $^{40}$Ca$(d,p)^{41}$Ca$_{\rm gs}$ and $^{48}$Ca$(d,p)^{49}$Ca$_{\rm gs}$ at $E_d{=}10$~MeV from the \emph{ab initio} CC+GFT framework of \cite{rotureauMergingInitioTheory2020}. 
  Curves correspond to different model--space truncations $N_{\max}$; experimental data are shown as filled circles. We also indicate the neutron separation energies resulting from the corresponding calculations. 
  The calculation associated with the larger largest space ($N_2{=}14$, $N_3{=}16$, where $N_2,N_3$ are the $N_{\text{max}}$ associated with 2- and 3-body forces, respectively) reproduces the angular distributions with high accuracy. When the neutron separation energy is adjusted to the experimental values ($S_n=-8.363$ MeV for $^{41}$Ca, $S_n=-5.146$ MeV for $^{49}$Ca) the absolute value of the cross section also compares very well with the experimental measurement. This last point illustrates how the GFT provides a natural normalization of the absolute value of the cross section without the need for spectroscopic factors (see Sect. \ref{CDWBA} and Fig. \ref{fig2}).}
  \label{fig:CaCCGFT}
\end{figure} 

\subsection{\texorpdfstring{$^{40,48}$Ca$(d,p)$}{x} with \emph{ab initio} CC+GFT}\label{SVB}

A central goal of modern nuclear theory is to connect nuclear reactions directly to the underlying nuclear interactions, without introducing ad hoc parameters. 
For decades, transfer reactions such as $(d,p)$ have been analyzed with the distorted--wave Born approximation (DWBA), which relies on phenomenological spectroscopic factors and adjustable optical potentials. 
By contrast, the recent work of Rotureau \emph{et al.}\  \cite{rotureauMergingInitioTheory2020}  demonstrates how to embed \emph{ab initio} structure information from the Coupled Cluster (CC) method into the Green's Function Transfer (GFT) framework. 
This unifies the description of structure and reactions: the overlaps, amplitudes and energies that govern the transfer are calculated consistently from the many--body Hamiltonian, here taken as NNLO$_\text{sat}$ chiral EFT including two-- and three--nucleon forces.

On the structure side, the CC approach provides ground--state energies and particle--addition amplitudes (PA--EOM) for closed-- and near--closed--shell systems. 
These overlap functions replace the spectroscopic factors of DWBA and directly enter the kernel of the GFT equations. 
On the reaction side, the GFT framework (see Sec.~\ref{Sec2}) expresses the $(d,p)$ cross section in terms of the single--particle Green's function of the neutron coupled to the target plus optical potentials for deuteron and proton scattering. 
The crucial point is that the same Hamiltonian---and in particular the same three--body forces---determine both the overlaps and the energies, so that the calculation has \emph{predictive power} once the interaction is fixed. 
The only phenomenological input is the choice of nucleon--target optical potentials, taken from global systematics when data are lacking.

The benchmark cases studied in \cite{rotureauMergingInitioTheory2020} are the ground--state to ground--state transfers $^{40}$Ca$(d,p)^{41}$Ca and $^{48}$Ca$(d,p)^{49}$Ca at $E_d=10$~MeV. 
For these stable targets, elastic scattering data exist and were used to constrain the optical potentials. 
Figure~\ref{fig:CaCCGFT} displays the resulting angular distributions. 
The curves show the CC+GFT predictions for increasing model--space truncations $N_{\max}$. 
The convergence pattern is somewhat non--monotonic, but the largest space ($N_2{=}14$, $N_3{=}16$) yields angular distributions in very good agreement with the experimental data, without adjustable scaling. 
The calculation slightly underbinds the final ground states ($\sim 0.5$--$0.6$~MeV), an effect corrected by shifting the energies to the experimental values while keeping the overlaps unchanged. 
This procedure corresponds to adjusting the proton momentum $k_p$ in the transfer kernel, not to refitting any spectroscopic parameters.

Once benchmarked against $^{40,48}$Ca, the method was applied to the exotic isotopes $^{52}$Ca and $^{55}$Ca, where no $(d,p)$ data exist. 
In these cases, optical potentials were taken from global parametrizations, while the overlaps and energies were still provided by CC. 
The calculations predict absolute cross sections and angular distributions, offering guidance for future experiments. 
The emergence of shell evolution with neutron number in the Ca chain, and the role of three--nucleon forces in shifting single--particle energies, are directly visible in the predicted transfer observables.

The Ca case study illustrates how \emph{ab initio} nuclear structure and reaction theories can be merged. 
By supplying overlaps and energies from CC into the GFT formalism, one obtains parameter--free predictions of $(d,p)$ angular distributions that agree with data in stable systems and extend to exotic nuclei where no measurements exist. 
This represents a significant step toward a unified, microscopic description of nuclear reactions based on chiral EFT interactions.

\section{Exclusive inelastic scattering}\label{SVI}
Eq. (\ref{eq409}) can be generalized to the calculation of the cross section for the population of a specific final state $i\neq0$ in the residual nucleus. This type of experiments are known as  \emph{exclusive} measurements, as opposed to the \emph{inclusive} ones discussed in Sect. \ref{Sec2}, where the final state is not identified and the cross section described is the \emph{total} reaction one, i.e., summed for all energy-accessible final states. The corresponding cross section is obtained   by projecting the wavefunction (\ref{eq557})  onto the specific final state $i$,
\begin{align}
 \nonumber \psi_{i}^{I}(\mathbf r_{xA})&=\int \Phi_{i}(\xi)\,\chi_{b}(\mathbf r_{bB},\mathbf k_{b})\Psi(\mathbf r_{xA},\mathbf r_{bB},\xi)\,d^3r_{bB}\,d\xi\equiv\bra{\Phi_{i}\,\chi_{b}(\mathbf k_{b})}\Psi\\
 \nonumber &=\bra{\Phi_{i}}\mathbf G(E-E_{b})\bra{\chi_{b}}\left[V_{xA}(\mathbf r_{xA},\xi)+U_{bA}(\mathbf r_{bA})\right]\Psi_{0}\\
 &=\int \Phi_{i}(\xi)\,\mathbf G(\mathbf r_{xA},\xi,E-E_{b})\chi_{b}(\mathbf r_{bB})\left[V_{xA}(\mathbf r_{xA},\xi)+U_{bA}(\mathbf r_{bA})\right]\Psi_{0}(\mathbf r_{xA},\mathbf r_{bB},\xi)\,d^3r_{bB}\,d\xi.
 \end{align}
Let us write the asymptotic wavefunction,
\begin{align}
 \psi_{i}^{I}(\mathbf r_{xA}\to \infty)=O(\mathbf k_{i},\mathbf r_{xA})\braket{\psi_{i}\Phi_{i}|V_{xA}(\mathbf r_{xA},\xi)|\Phi_{0}\psi^{HM}}=O(\mathbf k_{i},\mathbf r_{xA})\,T_{i0}^{I},
\end{align}
where we have used (\ref{eq410}) and the fact that 
\begin{align}
\braket{\psi_{i}\,\chi_{b}\Phi_{i}|U_{bA}|\Phi_{0}\phi_{a}F}=0,
\end{align}
and we have introduced the \emph{indirect} $T$-matrix,
\begin{align}\label{eq560}
T_{i0}^{I}=\braket{\psi_{i}\Phi_{i}|V_{xA}(\mathbf r_{xA},\xi)|\Phi_{0}\psi^{HM}},
\end{align}
which differs from the \emph{direct} $T$-matrix (\ref{eq404}) by the presence of $\psi^{HM}(\mathbf r_{xA})$ instead of the free wave $F(\mathbf r_{xA};\mathbf k_{xA})$ in the \emph{ket}.

 Let us introduce here a word of caution. Since an expression similar to Eq. (\ref{eq406}) cannot be established here, we have, in general, that
\begin{align}\label{eq432}
\braket{\psi_{i}\Phi_{i}|V_{xA}(\mathbf r_{xA},\xi)|\Phi_{0}\psi^{HM}}\neq \braket{\psi_{i}|\mathcal V|\psi^{HM}}.
\end{align}
This is a consequence of the fact that the \emph{optical reduction} process described in Sect. \ref{SOP} consists essentially in restricting the many-body operator $V_{xA}$ to the one-dimensional sector of the Hilbert space spanned by the ground state $\Phi_0$ of the nucleus $A$. Since the excited state $\Phi_i$ is outside this sector, the operator $V_{xA}$ cannot be reduced in Eq. (\ref{eq560}) to a single-particle operator $\mathcal V$. Although a generalization of the concept of optical potential by defining it in the space of all open reaction channels can be made (see, e.g., \cite{feshbachUnifiedTheoryNuclear1962,feshbachUnifiedTheoryNuclear1958}), we will not touch upon this subject here. Within this context, we do not propose in this Section a calculable method to obtain (\ref{eq560}). Instead, the main result of this Section is the explicit connection between the \emph{direct} and \emph{indirect} $T$-matrices expressed in Eq. (\ref{eq433}), and, more particularly, the relationship between the corresponding $R$-matrix parameters shown in Eq. (\ref{eq411}).

In order to establish a more explicit connection between the indirect and direct $T$-matrices, we will expand the Hussein-McVoy function in terms of a superposition of free waves,
\begin{align}
\psi^{HM}(\mathbf r_{xA})=\int g(\mathbf k)\,F(\mathbf r_{xA},\mathbf k)\,d\mathbf k,
\end{align}
where we have introduced a ``broadening factor'' $g$,
\begin{align}\label{eq501}
g(\mathbf k)=\int \psi^{HM}(\mathbf r_{xA})\,F^{*}(\mathbf r_{xA},\mathbf k)\,d\mathbf r_{xA}.
\end{align}
If $F$ is a plane wave (i.e., the fragment $x$ is a neutron, like in \cite{potelEstablishingTheoryDeuteroninduced2015}), the broadening factor $g$ is just the Fourier transform of the Hussein-McVoy function. 
In any case, it can be calculated if the ground state wavefunction $\phi_{a}(\mathbf r_{xb})$ of the nucleus $a$ is known which is, at any rate, a necessary prerequisite for the applicability of the indirect  method.  We can now write the indirect $T$-matrix in terms of the direct one,
\begin{align}\label{eq433}
T_{i0}^{I}=\int g(\mathbf k) T_{i0}(\mathbf k)\,d\mathbf k,
\end{align}
where
\begin{align}
T_{i0}(\mathbf k)=\braket{\psi_{i}\Phi_{i}|V_{xA}(\mathbf r_{xA},\xi)|\Phi_{0}F(\mathbf r_{xA},\mathbf k)}.
\end{align}
We now take advantage of the $R$-matrix parametrization (\ref{eq415}),
 \begin{align}
 T_{i0}(\mathbf k)=\sqrt{P_{i}(E_k)P_{0}(E_k)}\sum_{pq}\gamma_{ip}\,[A^{-1}(E_k)]_{pq}\,\gamma_{0q},
 \end{align}
with
\begin{align}
E_{k}=\frac{\hbar^{2}k^{2}}{2\mu}.
\end{align}
We can then write the explicit expression of the \emph{indirect} $T$-matrix in terms of the \emph{direct} $T$-matrix parameters,
 \begin{align}\label{eq411}
 T_{i0}^{I}=\int \sqrt{P_{i}(E_k)P_{0}(E_k)}\sum_{pq}\gamma_{ip}\,[A^{-1}(E_k)]_{pq}\,\gamma_{0q}\,g(\mathbf k)\,d\mathbf k.
 \end{align}
 \subsection{Discussion}
  The expression (\ref{eq411}) shows how the partial widths $\gamma_{ip},\gamma_{0q}$, as well as the energies $E_{p}$, can be fitted from the indirect experimental cross section and used to predict the direct one. 
  
 While the GFT expression (\ref{eq412}) can be used to \emph{calculate} the reaction cross section, Eq. (\ref{eq411}) suggests a \emph{fitting} procedure to the observed indirect cross section, proportional to $|T_{i0}|^{2}$. This fitting procedure is so similar in spirit to the $R$-matrix fit, that standard existing $R$-matrix codes (like, e.g., AZURE) might possibly be used without essential modification. The difference between the direct and indirect $R$-matrix parametrizations resides in the broadening factor $g$,
  \begin{align}
\nonumber g(\mathbf k)&=\int \psi^{HM}(\mathbf r_{xA})\,F^{*}(\mathbf r_{xA},\mathbf k)\,d\mathbf r_{xA}\\
&=\int \chi^{*}_{b}(\mathbf r_{bB},\mathbf k_{b})\phi_{a}(\mathbf r_{xb})F(\mathbf r_{aA},\mathbf k_{aA})\, F^{*}(\mathbf r_{xA},\mathbf k)\,d\mathbf r_{xA}\,d\mathbf r_{xb}.
\end{align}
 The form of the above integral suggests that $g(\mathbf k)$ is peaked around $\mathbf k=\mathbf k_{aA}-\mathbf k_{b}$, and has a width of the order of the momentum spread of the Fourier transform of $\phi_{a}$. Qualitatively,
 \begin{align}
 g(\mathbf k)\sim \widetilde \phi_{a}(\mathbf k_{aA}-\mathbf k_{b}),
 \end{align}
where $\widetilde \phi_{a}$ is the Fourier transform of $\phi_{a}$. 
If the shift and penetrability factors vary slowly in such an interval, we might be able to approximate
 \begin{align}\label{eq418}
 \nonumber T_{i0}^{I}\approx& \sqrt{P_{i}(E_{k_0})P_{0}(E_{k_0})}\sum_{pq}\gamma_{ip}\,[A^{-1}(E_{k_0})]_{pq}\,\gamma_{0q}\int \,g(\mathbf k)\,d\mathbf k\\
 &=N\,\sqrt{P_{i}(E_{k_0})P_{0}(E_{k_0})}\sum_{pq}\gamma_{ip}\,[A^{-1}(E_{k_0})]_{pq}\,\gamma_{0q},
 \end{align}
where $E_{k_{0}}$ is the energy at the peak of the momentum distribution, and the constant $N$ is
\begin{align}
N=\int \,g(\mathbf k)\,d\mathbf k.
\end{align}
Eq. (\ref{eq418}) has the form suggested by Barker \cite{barkerInterferenceLevelsNuclear1967} on what seem to be purely heuristic grounds, where the role of what he calls the ``feeding factor'' $G_{abq}$ is played here by 
\begin{align}
G_{abq}=\gamma_{0q}N.
\end{align}
We haven't been able to find in \cite{barkerInterferenceLevelsNuclear1967} or elsewhere any expression or derivation of this feeding factor $G_{abq}$. This approximation ignores completely the energy-dependence of the $T$-matrix over the distribution defined by the broadening factor $g$. It should also be pointed out that the modification of the entrance channel introduced by the broadening factor is also angular momentum dependent, and can be calculated from the multipolar expansion of $g(\mathbf k)$. In any case, the validity of these approximations can be checked by directly computing $g(\mathbf k)$, and using the exact expression (\ref{eq411}). 

A slightly better approximation could be to account for the additional energy width $\Delta E$ introduced by the broadening phenomenon by adding a corresponding imaginary part in the denominator,
 \begin{align}\label{eq414}
 T_{i0}^{I}\approx N\,\sqrt{P_{i}(\tilde E)P_{0}(\tilde E)}\sum_{pq}\gamma_{ip}\,[A^{-1}(\tilde E)]_{pq}\,\gamma_{0q},
 \end{align}
 with $\tilde E=E_{k_{0}}+i\Delta E$. This illustrates the fact that the factor $g$ broadens the resonances observed with the indirect method with respect to their ``direct values''. 
\section{Beyond the spectator approximation}\label{SVIB}
Within the spectator approximation, the dynamical role of the cluster $b$ is essentially  only to ``carry''  the cluster $x$ inside the projectile $a$, without altering the way in which $x$ and $A$ interact with each other. Within this context, the un-scattered  state in the $x-A$ collision doesn't correspond, like in the direct measurement of the $x-A$ scattering process, to a state of well defined energy described by a free wave $F$, but is instead modified by the intrinsic motion of $x$ in the ground state of the projectile $a$, as testified by the broadening factor $g$ (\ref{eq501}). In other words, the role of the un-scattered wavefunction is now played by the Hussein-Mc Voy term (\ref{eq:HM}), as it becomes apparent by comparing Eqs. (\ref{eq317}) and (\ref{eq325}). Within this approximation, the explicit coupling of $b$ with excited states of $A$ or $B\equiv A+x$ is completely ignored. In order to gain some insight concerning the validity of this approximation, we need to go beyond the spectator approximation. 

Let us start by writing the \emph{exact} version of Eqs. (\ref{eq551}) and (\ref{eq408}),
\begin{align}\label{eq552}
\left(E-T_a-h_a(\mathbf r_{xb})-h_A(\xi)\right)\Psi=\left(V_{xA}(\mathbf r_{xA},\xi)+U_{bA}(\mathbf r_{bA})+\Delta V_{A}\right)\Psi,
\end{align}
and
\begin{align}\label{eq553}
\Psi=\Psi_{0}+\mathcal G(E)\left[V_{xA}(\mathbf r_{xA},\xi)+U_{bA}(\mathbf r_{bA})+\Delta V_{A}\right]\Psi_{0},
\end{align}
Where $\Delta V_{A},\Delta V_{B}$ are defined in Eq. (\ref{eq328}), and  the \emph{exact} Green's function $\mathcal G(E)$ can be expressed in terms of the spectator approximation Green's function,
\begin{align}\label{eq554}
\nonumber \mathcal G(E)&=\mathbf G_{I}(E)+\mathbf G_{I}(E)\,\Delta V_{B}\,\mathcal G(E)= \mathbf G_{I}(E)+\mathbf G_{I}(E)\,\Delta V_{B}\,\mathbf G_{I}(E)\\
&+(\text{terms of higher order in }\Delta V_B),
\end{align}
From the two expressions above we obtain the exact wavefunction, 
\begin{align}\label{eq555}
 \Psi= \Psi_{0}+\Psi_{xA}+\mathbf G_{I}(E)\,\Delta V_{A}\,\Psi_{0}+\mathbf G_{I}(E)\,\Delta V_{B}\,\Psi_{xA}+(\text{terms of higher order in }\Delta V_A,\,\Delta V_B),
\end{align}
where the two first terms in the right hand side correspond to the spectator approximation (\ref{eq557}), including the scattered wavefunction $\Psi_{xA}$, while the 3rd and 4rth terms provide the corrections to first order in $\Delta V_A$ and $\Delta V_B$. Let us further assume that  
\begin{align}\label{eq326}
 \Psi_{0}\gg\Psi_{xA},
\end{align}
i.e., that the departure away from the ``elastic'' channel represented by the scattered wave $\Psi_{xA}$ is small. This seems all the more reasonable if we remember that one can (and actually, most of the time, does) implement the distorted wave strategy (see Sect. \ref{DWBA} and, in particular, Eq. (\ref{eq321})), without essentially altering the present discussion. Then the wavefunction corresponding to the spectator approximation plus the first order correction is,
\begin{align}\label{eq327}
 \Psi\approx \Psi_{0}+\Psi_{xA}+\mathbf G_{I}(E)\,\Delta V_{A}\,\Psi_{0}=\Psi_{0}+\Psi_{xA}+\mathbf G_{I}(E)\,(V_{bA}(\mathbf r_{bA},\xi)-U_{bA}(\mathbf r_{bA}))\,\Psi_{0}.
\end{align}
In order to describe the elastic process associated with the detection of particle $b$, we proceed as for the derivation of Eq. (\ref{eq409}) projecting onto the state $\Phi_{0}\chi_{b}(\mathbf k_{b})$,
\begin{align}\label{eq556}
 \nonumber  \bra{\Phi_{0}\,\chi_{b}(\mathbf k_{b})}\Psi&\approx\psi^I_{1}=\psi^{HM}+\bra{\Phi_{0}}\mathbf G(E-E_{b})\bra{\chi_{b}}\left [V_{xA}(\mathbf r_{xA},\xi)+U_{bA}(\mathbf r_{bA})\right ]\Psi_{0}\\
 \nonumber &+\bra{\Phi_{0}}\mathbf G(E-E_{b})\bra{\chi_{b}}\,\left(V_{bA}(\mathbf r_{bA},\xi)-U_{bA}(\mathbf r_{bA})\right)\,\Psi_{0}\\
 &=\psi_{0}^{I}(\mathbf r_{xA})+\bra{\Phi_{0}}\mathbf G(E-E_{b})\bra{\chi_{b}}\,\left(V_{bA}(\mathbf r_{bA},\xi)-U_{bA}(\mathbf r_{bA})\right)\Psi_{0},
\end{align}
where we have used Eq. (\ref{eq325}) to identify the  spectator approximation wavefunction $\psi_{0}^{I}$. In order to get some insight into the meaning of the correction to the spectator approximation, let us use the explicit expressions of $\mathbf G$ and $\Psi_0$ (see Eqs. (\ref{eq329}) and (\ref{eq330})),
\begin{align}\label{eq331}
 \nonumber \mathbf G(E)&= \sum_{ij}\ket{\Phi_{i}(\xi)}\left(\left(E_{i}-E_{b}-T_{x}(\mathbf{r}_{xA}))\right)\delta_{ij}-V_{ij}(\mathbf{r}_{xA})\right)^{-1}\bra{\Phi_{j}(\xi)};\\
  \ket{\Psi_{0}}&=\ket{\phi_{a}(\mathbf r_{xb})\Phi_{0}(\xi)F (\mathbf k_{a},\mathbf r_{a})}.
\end{align}
Substituting in (\ref{eq556}), the corrected wavefunction can be written as
\begin{align}\label{eq558}
 \nonumber  \psi^I_{1}&=\psi_{0}^{I}(\mathbf r_{xA})
  +\sum_{i}\left(\left(E_{i}-E_{b}-T_{x}(\mathbf{r}_{xA}))\right)\delta_{0i}-V_{0i}(\mathbf{r}_{xA})\right)^{-1}\\
  \nonumber &\times \bra{\Phi_{i}\chi_{b}}\,\left(V_{bA}(\mathbf r_{bA},\xi)-U_{bA}(\mathbf r_{bA})\right)\ket{\phi_{a}\Phi_{0}F}\\
   \nonumber &=\psi_{0}^{I}(\mathbf r_{xA})
  +\sum_{i}\left(\left(E_{i}-E_{b}-T_{x}(\mathbf{r}_{xA}))\right)\delta_{0i}-V_{0i}(\mathbf{r}_{xA})\right)^{-1}\\
  &\times\left[\bra{\Phi_{i}\chi_{b}}\,V_{bA}(\mathbf r_{bA},\xi)\ket{\phi_{a}\Phi_{0}F)}-\bra{\chi_{b}}U_{bA}(\mathbf r_{bA})\ket{\phi_{a}F}\delta_{0i}\right].
\end{align} 
We will not try here to give a quantitative estimate of the importance of the correction to the spectator approximation contained in the 2nd term in the right hand side of he above expression. However, it is still possible to extract some physical insight from it, which will help assessing the regime of validity of the spectator approximation.

The terms with $i\neq0$ represent the inelastic excitations of $A$ due to the interaction of $b$ with $A$, and they contribute to the scattered wave with an amplitude proportional to the matrix elements
\begin{align}
\bra{\Phi_{i\neq0}\chi_{b}}\,V_{bA}(\mathbf r_{bA},\xi)\ket{\phi_{a}\Phi_{0}F)}.
\end{align}
Part of this contribution is compensated by the removal of flux associated with the imaginary part of the optical potential $U_{bA}$. However, standard optical potentials with a smooth energy dependence are associated with an energy-averaged description of the interaction of $b$ with $A$. Therefore, they do not account for the contribution of narrow, tightly spaced resonances of the $b+A$ system present at low energies, which give rise to a rapidly varying energy dependence of the scattering amplitude. The difference between the energy-averaged cross section accounted for by standard optical potentials and the rapidly-varying one is called the \emph{fluctuation} cross section, and is characteristic of compound nuclear reactions \cite{FriedmanWeisskopf1955}.

The term corresponding to $i=0$ also contributes to the correction term, with an amplitude proportional to
\begin{align}\label{eq333}
\bra{\Phi_{0}\chi_{b}}\,V_{bA}(\mathbf r_{bA},\xi)\ket{\phi_{a}\Phi_{0}F)}-\bra{\chi_{b}}U_{bA}(\mathbf r_{bA})\ket{\phi_{a}F}.
\end{align}
This term represents the contribution associated with elastic processes not accounted for by the optical potential $U_{bA}$. As for the inelastic processes discussed above, the origin of this difference lies in the contribution of narrow resonances of the $b+A$ system which, in this case, decay by emitting $b$ back into the elastic channel. In this context, the fluctuation cross section is called the \emph{compound elastic} contribution \cite{FriedmanWeisskopf1955}.

This discussion  suggests that the spectator approximation might break whenever the incident effective energy of $b$, determined by the peak of the broadening factor $g(\mathbf k_b)$ (see Eq. (\ref{eq501})), is in the region of low-energy resonances of the $b+A$ system. Therefore, in order to take advantage of the spectator approximation one should devise experiments with a high enough incident energy of the projectile $a$. The same considerations apply to the exit channel, suggesting that the selected events should correspond to a high enough energy of the detected particle $b$.

A possible physical picture that emerges from the structure of (\ref{eq555}) and this discussion is the following. The spectator approximation accounts exactly for the scattering of $x$ with $A$, while the interaction of $b$ with $A$ is treated in an effective way through the optical potential $U_{bA}$. The resulting scattered wave is $\Psi_{xA}$. The first correction (third term in (\ref{eq555})) consists in accounting for the scattering of $b$ with $A$ to first order in the difference between the real interaction $V_{bA}$ and the optical potential $U_{bA}$. This correction accounts for the excitation of $A$ by $b$ and for elastic processes not accounted for by the optical potential $U_{bA}$. Up to now, we have considered single-scattering events, while the next correction (fourth term in \ref{eq555}) accounts for a double-scattering process: first, $x$ scatters with $A$, and then the interaction $\Delta V_B=V_{xb}(\mathbf r_{xb})+V_{bA}(\mathbf r_{bA},\xi)-U_{bB}(\mathbf r_{bB})$ drives a simultaneous scattering event of  $b$  with $A$ and $x$. This is the lowest order at which the genuinely three-body character of the system appears, arising from the three-body nature of the interaction $\Delta V_B$. 

Let us finally emphasize again that the soundness of this perturbative scheme rely on $\Delta V_A$ and $\Delta V_B$ being small. More specifically, the matrix elements of the kind of those appearing in (\ref{eq333}) should be small. We have tried to show here that this rely on two basic assumptions: (i) the optical potential $U_{bA}$ should provide a good energy-averaged description of the interaction of $b$ with $A$, and (ii) the incident energy of $a$, and the final energy of $b$, should be high enough to avoid the region of low-energy resonances of the $A$ and $x+A$ systems, where standard optical potentials fail to describe the fluctuation cross section.
\subsection*{Acknowledgments}
I wish to thank E. Vigezzi and F. Barranco for a critical reading of the manuscript and for many insightful suggestions.

\end{document}